\documentclass[a4paper,fleqn,usenatbib]{mnras}

\usepackage[T1]{fontenc}
\usepackage{ae,aecompl}

\usepackage{graphicx}	% Including figure files
\usepackage{amsmath}	% Advanced maths commands
\usepackage{amssymb}	% Extra maths symbols
\usepackage{comment}
\usepackage{xcolor}
\usepackage{enumitem}
\usepackage{soul}
\usepackage{hyperref}
\definecolor{temporary}{rgb}{0.7, 0.7, 0.01}

\definecolor{seagreen}{rgb}{0.190, 0.525, 0.361}
\definecolor{steelblue}{rgb}{0.275, 0.51, 0.706}

\newcommand{\micmap}[1]{\textcolor{red}{#1}}
\newcommand{\AB}[1]{\textcolor{magenta}{#1}}
\newcommand{\GN}[1]{\textcolor{purple}{#1}}
\newcommand{\ST}[1]{\textcolor{seagreen!100}{#1}}
\newcommand{\NGsp}[1]{\textcolor{orange}{#1}}
\newcommand{\MP}[1]{\textcolor{steelblue!100}{{#1}$_{\mathrm{MP}}$}}
\title[Binaries in young stellar clusters]{The impact of binaries on the evolution of star clusters from turbulent molecular clouds}

\author[S. Torniamenti et al.]{Stefano Torniamenti$^{1,2,3}$\thanks{E-mail: stefano.torniamenti@studenti.unipd.it}, 
Alessandro Ballone$^{1,2,3}$,    
Michela Mapelli$^{1,2,3}$\thanks{E-mail: michela.mapelli@unipd.it}, 
Nicola Gaspari$^{1,4}$,
\newauthor Ugo N. Di Carlo$^{1,2,3,5}$, 
Sara Rastello$^{1,2}$, 
Nicola Giacobbo$^{1,2,6}$,
Mario Pasquato$^{7}$
\\
$^{1}$Physics and Astronomy Department Galileo Galilei, University of Padova, Vicolo dell'Osservatorio 3, I--35122, Padova, Italy\\
$^{2}$INFN - Padova, Via Marzolo 8, I--35131 Padova, Italy\\
$^{3}$INAF - Osservatorio Astronomico di Padova, Vicolo dell'Osservatorio 5, I-35122 Padova, Italy\\
$^{4}$Department of Astrophysics/IMAPP, Radboud University, P.O. Box 9010, 6500 GL, Nijmegen, The Netherlands\\
$^{5}$Dipartimento di Scienza e Alta Tecnologia, University of Insubria, Via Valleggio 11, I–22100, Como, Italy \\
$^{6}$School of Physics and Astronomy, Institute for Gravitational Wave Astronomy, University of Birmingham, Birmingham, B15 2TT, UK \\
$^{7}$Center for Astro, Particle and Planetary Physics (CAP3), New York University, Abu Dhabi
}

\date{Accepted XXX. Received YYY; in original form ZZZ}

\pubyear{2021}

\begin{document}
\label{firstpage}
\pagerange{\pageref{firstpage}--\pageref{lastpage}}
\maketitle

% Abstract of the paper
\begin{abstract}
Most of massive stars form in binary or higher-order systems in clumpy, sub-structured clusters. In the very first phases of their life, these stars are expected to interact with the surrounding environment, before being released to the field when the cluster is tidally disrupted by the host galaxy. 
%Modelling this complex interplay is of fundamental importance to understand the properties of the observed binary populations. 
We present a set of \textit{N}$-$body simulations to describe the evolution of young stellar clusters and their binary content in the first phases of their life. To do this, we have developed a method that generates realistic initial conditions for binary stars in star clusters from hydrodynamical simulations. We considered different evolutionary cases to quantify the impact of binary and stellar evolution. Also, we compared their evolution to that of King and fractal models with different length scales.
Our results indicate that the global expansion of the cluster from hydrodynamical simulations is initially balanced by the sub-clump motion and  accelerates when a monolithic shape is reached, as in a post-core collapse evolution. 
%Also, 
Compared to the spherical initial conditions, the ratio of the 50\% to 10\% Lagrangian radius  %$r_{50}/r_{10}$ 
shows a very distinctive trend, explained by the formation of a hot core of massive stars triggered by the high initial degree of  mass segregation. 
As for its binary population, each cluster shows a self-regulating behaviour by creating interacting binaries with binding energies of the order of its energy scales. Also, in absence of original binaries, the dynamically formed binaries present a mass dependent binary fraction, that mimics the trend of the observed one.

\end{abstract}

% Select between one and six entries from the list of approved keywords.
% Don't make up new ones.

%%%%%%%%%%%%%%%%%%%%%%%%%%%%%%%%%%%%%%%%%%%%%%%%%%

%%%%%%%%%%%%%%%%% BODY OF PAPER %%%%%%%%%%%%%%%%%%

\section{Introduction}\label{intro}

Most stars form as members of clusters  or associations, that present a clumpy spatial distribution and may also contain sub-structures (\citealp{larson95}). %\AB{\textbf{AB: Non mi convince questa prima frase... Non direi che la distribuzione di stelle sia turbolenta... Poi, qualcosa di clumpy contiene sotto-strutture per definizione...}} 
Understanding the early evolution of these star-forming complexes is of fundamental importance for the comprehension of the properties of young ($<100$ Myr) 
%\AB{\textbf{AB: young sono di solito meno vecchi di 100 Myr... dove hai trovato questa definizione?}} 
and open clusters \citep{Portegies-Zwart10}, where the presence of sub-structures and fractality is observed \citep[e.g.,][]{Cartwright04,Sanchez09,Parker12,Kuhn19}. Also, these systems are characterized by complex internal kinematics, such as  sub-clump relative motions and mergers, cluster expansion, gas dispersal (\citealp{Kuhn19,Cantat-Gaudin19}) and rotation (\citealp{Henault-Brunet12}).  
In particular, gas  dispersal due to stellar winds and supernova explosions drives the cluster out of dynamical equilibrium, %generating a cluster expansion 
leading to an expansion phase, where most stars become unbound and disperse into the field \citep{hills80,goodwin06,baumgardt07,pfalzner09}. 
Some of these natal properties might even survive the successive evolution of the stellar system and leave an imprint on the observed properties of older, relaxed stellar clusters (e.g., they may contribute to the signatures of rotation visible in some globular clusters, \citealp{vanLeeuwen00,Pancino07,Bianchini13,Kamann18,Bianchini18}). 
%\micmap{MM: Qui rischi il backfire: non \`e detto che la rotazione che vediamo nei globular cluster sia pristina. ST: ho cercato di aggiustare.}

In the first phases of their life, the dynamical evolution of young stellar clusters is deeply influenced by their stellar and binary content, and vice versa. In particular, a large fraction of the most massive stars is %observed to be 
part of binary and higher order systems \citep{moe17} that can actively exchange energy and angular momentum with the host environment, thanks to the very high density  ($\rho \sim 10^4$ M$_{\odot}$ pc$^{-3}$) of the cluster core. 
On the one hand, original binary stars (i.e., stars that form as members of a binary system) contain a large reservoir of internal energy, that can be transferred to other stars in the host star cluster, through three- and multi-body encounters \citep[e.g.,][]{heggie75,hut83}, preventing or reversing the gravothermal collapse of the core of the cluster \citep{chatterjee13,fujii14}. 
On the other hand, the global evolution of the cluster affects the properties of the binary population: for example, core collapse leads to the formation of new binary systems and to their dynamical hardening \citep{spitzer71}. 
On top of this, binary stars are also affected by mass transfer, common envelope, supernova kicks, tides and other evolutionary processes \citep[e.g.,] []{hut81,webbink84,portegieszwart96,hurley02}. All these processes are crucial for the ejection of stars from their host star cluster (e.g., runaway stars, \citealp{fujii11}), and for the formation of intermediate-mass black holes \citep[e.g.,][]{Ebisuzaki01,portegieszwart04,Giersz15,mapelli16}. Finally, the interplay between dynamical interactions and binary evolution \citep{banerjee10,ziosi14,banerjee17,fujii17,dicarlo20,kumamoto19,antonini20,trani21} can explain the properties of the binary compact objects observed through gravitational wave detection by LIGO and Virgo \citep{abbottO3a,abbottO3apopandrate}. 

Direct \textit{N}$-$body simulations are usually adopted to integrate the collisional dynamics of gas-free star clusters, where length-scales of different orders of magnitude, from binary separations of some solar radii to several parsecs, need to be included. However, studies of this type often lack realistic initial conditions. 
%and/or a self consistent stellar mass-function and binary population.
For example, state-of-the-art direct \textit{N}$-$body simulations of star clusters include realistic stellar mass functions and stellar evolution, but most of them start from spherical idealized models, such as \cite{plummer11} or  \citet{king66} models.  
In some recent work, fractal initial conditions were adopted to mimic the initial clumpiness of star clusters \citep[e.g.,][]{Goodwin04,Schmeja06,Allison10,Kupper11,Parker14b,dicarlo19,Daffern-Powell20}. Few studies tried to re-simulate with a direct \textit{N}$-$body code the initial conditions obtained from hydrodynamical simulations of star cluster formation \citep{Moeckel10,Moeckel12,Parker13,Fujii15a}, but most of them do not include stellar evolution or realistic stellar mass functions or original binary populations. A recent attempt to couple magneto-hydrodynamics and direct \textit{N}$-$body star cluster formation simulations, also considering the presence of original binaries, was proposed by \cite{cournoyercloutier20}, who developed a binary generation algorithm consistent with observations of mass dependent binary fraction and distributions of orbital periods, mass ratios and eccentricities. They found that binary systems formed dynamically do not have the same properties as the original ones, and that the presence of an initial population of binaries affects the properties of dynamically-formed binaries. 
%with respect to the case with only single stars. 
An adequate modelling of the original binary population is thus necessary for a realistic description of dynamical interactions in %that take place during 
the early stages of star clusters' evolution. 

Recently, \cite{ballone20} and \cite{ballone21} proposed a new approach to connect hydrodynamics and stellar dynamics that can be used to provide more realistic initial conditions for direct \textit{N}$-$body simulations. This approach includes a number of the ingredients necessary to self-consistently study this problem: realistic phase-space distributions of stars, drawn from sink particle distributions of collapsing molecular clouds, and a realistic stellar mass function, which is fundamental to assess the impact of stellar evolution. This method is based on the assumption that the gas, in which the newly formed star cluster is embedded, is almost instantaneously expelled by feedback (radiation, winds and, most of all, supernova explosions) from the young most massive stars \citep[e.g.,][]{VazquezSemadeni10,Dale14,pfalzner14,gavagnin17,chevanche20b,chevanche20a,2020ApJ...900L...4P}. From that moment on, the evolution of the newly born stellar system is mainly driven by gravitational dynamics. A necessary step towards a more realistic description is the insertion of binary stars in the original stellar population. 

The aim of this paper is to offer a realistic, self-consistent description of the complex interplay between binaries and their host cluster in the first phases of a cluster's life after gas expulsion, by considering the effects of dynamics, stellar and binary evolution simultaneously. To do this, we %implement the insertion of 
insert original binaries in the joining/splitting method introduced in \citet{ballone21}, to generate realistic initial conditions for \textit{N}$-$body simulations starting from hydrodynamical simulations. % adopting a sink particle algorithm. 
Also, we study the evolution of the phase-space distribution of star clusters generated by hydrodynamical simulations and we compare it to other, more idealized, initial configurations.

This paper is organized as follows. In Sect.~\ref{sec_methods}, we introduce our binary generation algorithm. Section~\ref{sec_initial_coditions} describes the initial conditions of the \textit{N}$-$body simulations. In Sect.~\ref{sec_results}, we report the results of the simulation of a stellar cluster under different evolutionary conditions and compare it to other initial phase-space distributions. In Sect.~\ref{sec_discussion}, we discuss the peculiar aspects of the evolution of the stellar clusters from hydrodynamical simulations. Finally, in Sect.~\ref{sec_conclusions} we report our conclusions.

\section{Methods}\label{sec_methods}

%%%%%%%%%%%%%%%%%%FIGURE%%%%%%%%%%%%%%%%%%%%%%%%%%%%%%%%%
\begin{figure*}
\begin{center}
\includegraphics[width=\textwidth]{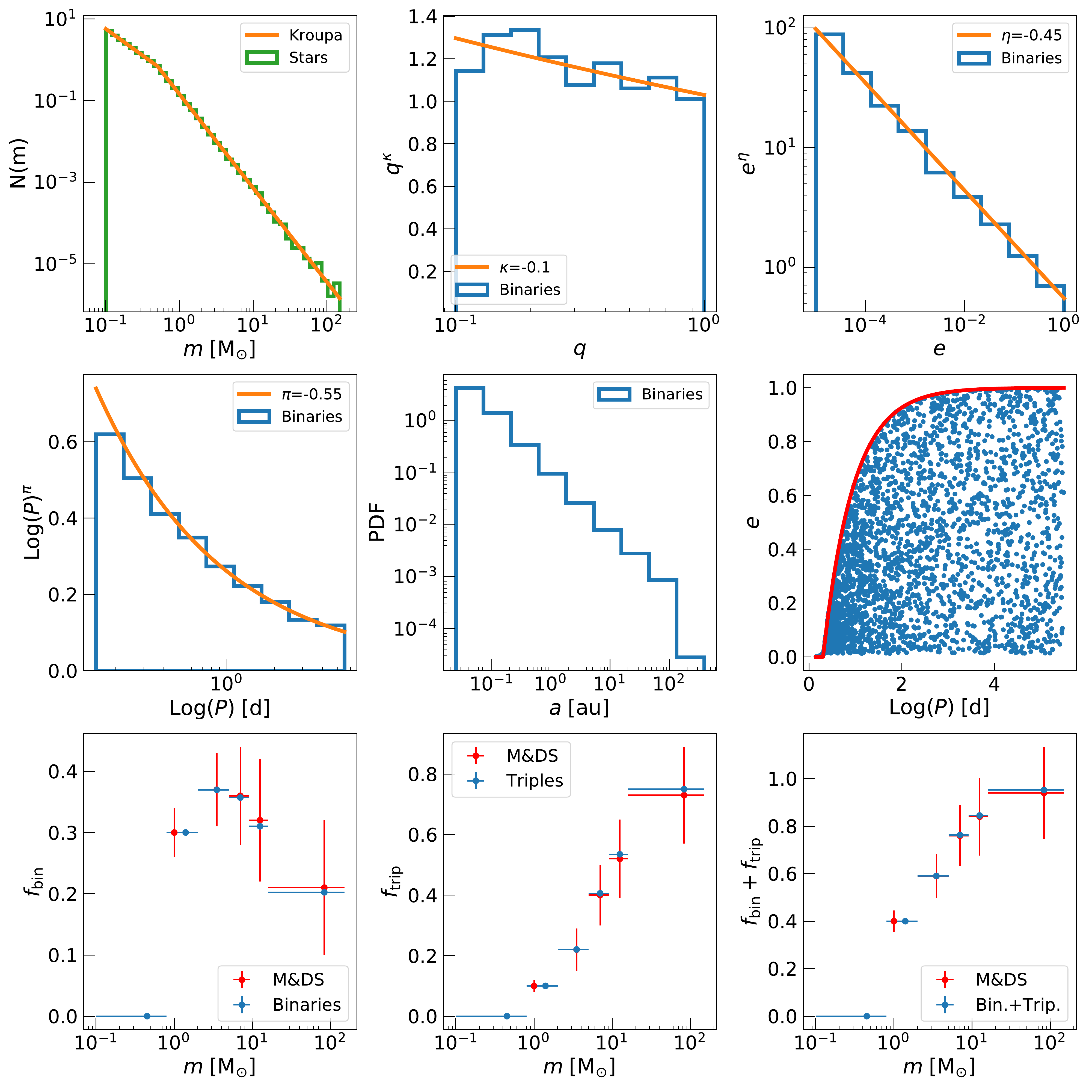}
\caption{Properties of binary stars generated with the algorithm described in Section\protect{~}\protect{\ref{sec_bin_gen}}. \textbf{Upper panel:} \protect\cite{Kroupa01} IMF (left), mass ratio (centre) and eccentricity (right) distributions, following \protect\cite{sana12}. \textbf{Central panel:} period distribution (left) from \protect\cite{sana12}, the resulting semi-major axis distribution (centre), and the eccentricity\protect{$-$}period relation (right) from \protect\cite{moe17}. \textbf{Lower panel:} fraction of binaries not counted as triples (\protect{$f_{\mathrm{bin}}$}, left), fraction of binary stars counted as triples (\protect{$f_{\mathrm{trip}}$}, centre) and the resulting binary fraction (\protect{$f_{\mathrm{bin}}+f_{\mathrm{trip}}$}, right). Red data points labelled as \protect{M\&DS}: observations from \protect\cite{moe17}. Blue data points: simulated binaries and triples from this work.}\label{fig_binary_distributions}
\end{center}
\end{figure*}
%%%%%%%%%%%%%%%%%%FIGURE%%%%%%%%%%%%%%%%%%%%%%%%%%%%%%%%%

\subsection{Binary generation algorithm} \label{sec_bin_gen}

We developed a new algorithm to generate a realistic initial mass function (IMF) and a realistic population of original binaries, based on  observations \citep{sana12,moe17}. This algorithm can be easily coupled to different phase-space generation codes to obtain a  variety of initial conditions for \textit{N}$-$body simulations. The method consists of the following steps.
\begin{enumerate}[label=\alph*),align=left]
    \item First, the algorithm randomly draws a population of stars from a \cite{Kroupa01} IMF between $0.1 \; M_{\odot}$ and $150 \; M_{\odot}$, for an assigned value of the total mass of the population.
    
    \item The stars are paired up to each other in order to obtain a distribution of mass ratios $q=m_2/m_1$ following \cite{sana12}:
    \begin{equation}
    \mathcal{F}(q) \propto q^{-0.1}, \;  \mathrm{with} \; q \; \in [0.1,\,{}1]. 
    \end{equation}
    The coupling is set to generate a binary fraction  $f_{\rm bin} = N_{\rm bin} / (N_{\rm sing} + N_{\rm bin})$, where $N_{\rm bin}$ is the number of binary systems and $N_{\rm sing}$ is the number of single stars, which depends on the mass of the primary star, following the observational results of \cite{moe17}. 
    For simplicity's sake, we do not include triple systems, but we take into account their presence when evaluating the binary fraction by labeling a certain number of single stars as third components of the existing binary systems (following \citealt{moe17}). This results in a fraction of binaries counted as triples ($f_{\mathrm{trip}}$), and prevents from having an excessive number of binary systems among the most massive stars. By this procedure, we obtain a distribution %of particles made 
    of single stars and of binary particles. %(whose mass is the sum of the masses of the binary components).
    For this work, we assume the binary fraction goes to zero in the mass range $0.1-0.8 \; M_{\odot}$: the observations indicate that the percentage of binary stars in this mass range is low anyway \citep{moe17}, and including these low-mass binary stars would have dramatically increased the computational cost of the simulations. 
    %did not consider any binary system in the mass bin $0.1-0.8 \; M_{\odot}$ because we are not interested in the behavior of the least massive binaries (\AB{\textbf{AB: dobbiamo spiegare perch\'e non ci interessa...}}) and they would have tremendously increased the computational cost of the simulations. 
   % The resulting binary fraction is $f_{\rm bin} = 0.06$. 
   The resulting binary fraction for stars with mass $m> 0.8 \; M_{\odot}$ is $0.4$. 
    
    \item The single and binary particles are assigned a phase-space distribution by coupling the aforementioned algorithm to a phase-space distribution generator. For this work, we considered two choices of the phase-space distribution generator. 
    In the first case, we coupled our algorithm with the joining/splitting procedure  summarized in the next sections and described in detail in \cite{ballone21}. %The joining/splitting procedure is not altered by the presence of binary particles, that are split or joined as ones. 
    %However, with respect to the splitting of single stars, more sink particles are needed to form the most massive binary particles (that have masses $>150 \; M_{\odot}$). This, in turn, induces more mass segregation in the final distribution of stars. 
    In the second case, the phase-space distribution  is created with the code {\sc McLuster} \citep{Kupper11}. 
    %\st{In particular, after running McLuster, the created binary stars are merged to obtain binary particles in analogy to those generated by our algorithm in the previous step. Then, we replace the original distribution of binary particles with the new one, starting from the most massive ones.} \micmap{Onestamente nella parte che ho stroked non si capisce proprio cosa stai dicendo. Ma \`e davvero necessario scendere in questi dettagli, tra l'altro spiegati male?} As this procedure may cause a variation in the virial state of the cluster, the velocities of the new distribution are rescaled to retain the same virial ratio as the original cluster.
    \item Finally, the binary particles are split into separate stars and their orbital period ($P$) and  eccentricity ($e$) distributions are generated following \cite{sana12}:  
    \begin{equation}
    \mathcal{F}(\mathcal{P}) \propto \mathcal{P}^{-0.55}, \; \mathrm{with} \; \mathcal{P}=\log_{10}(P/ {\rm days})\in [0.15,\,{}5.5],
    \end{equation}
    and 
    \begin{equation}
    \mathcal{F}(e) \propto e^{-0.45}, \; \mathrm{with} \; e \in [10^{-5},\,{}e_{\rm max} (P)],
    \end{equation}
    %The semi-major axis distribution is the obtained by applying the Third Kepler's law. 
    where, for a given orbital period, we set the upper limit of the eccentricity distribution $e_{\rm max} (P)$ according to eq.~(3) of \cite{moe17}: 
    \begin{equation}
        e_{\rm max} (P) = 1 - \left( \frac{P}{2 \, {\rm days}} \right)^{-2/3}. 
    \end{equation}
    The orbital properties of the binaries are then converted into positions and velocities by considering an isotropic  distribution for the orbital planes.
\end{enumerate}

\begin{comment}
\GN{Alcune curiosit\`a: 1) il check per vedere se le binare mergono subito era attivo? 2) usando eq. 4 significa che il range di eq 5 diventa [1e-5, $e_{\rm max}$(P)], giusto? 3) usando eq. 4 non dovremmo discorstarci leggermente da eq. 3?} \ST{1) No, il check lo abbiamo tolto quando abbiamo aggiunto la relazione tra eccentricità e periodo 2) sì, hai ragione, ho specificato meglio, grazie }
%3) Se consideriamo binarie con periodi log(P)>2 il limite non dovrebbe essere troppo diverso (almeno così sembra dal pannello centrale a destra di figura 1), per sistemi con periodo breve la differenza è più marcata.) 
\NGsp{NGsp: Due parole sul ruolo di eq4. Nella mia implementazione, dopo aver generato le masse delle binarie, le si conta e si generano altrettanti periodi ed eccentricità seguendo Sana+12. Poi si assegna un'eccentricità ad ogni binaria, e si pesca un periodo random tra quelli che soddisfano eq4. Quindi a (3) risponderei no.} \GN{X NGsp: continuo ad essere convinto che ci sia una (piccola) discrepanza (magari ne discuteremo meglio a voce :)). Graze ad entrambi.}
\end{comment}

Figure~\ref{fig_binary_distributions} shows an example of the binary populations generated by means of this algorithm. These initial conditions can be used to study the evolution of binary stars at the early and later stages of their host stellar cluster's life with a great variety of initial configurations. In addition, the generation of initial conditions through this algorithm has negligible computational cost. Finally, the described procedure is also suited to generate initial conditions for population synthesis studies.
\begin{comment}
\MP{Una domanda: gli altri elementi kepleriani? Fissata eccentricita, periodo ecc... mi rimane da decidere come si orienta il piano dell'orbita rispetto al cluster e alle altre binarie: sono tutte complanari? Oppure la direzione della velocita viene randomizzata? E poi partono tutte al pericentro o all'apocentro o con una fase casuale? Forse non serve dirlo ma ero curioso. Tra l'altro c'è evidenza che i piani delle binarie siano allineati nei cluster veri \citep[][]{2017NatAs...1E..64C}} 
\end{comment}

\subsection{Hydrodynamical simulations} \label{sec_hydro_sim}

%As discussed in Section \ref{sec_bin_gen}, %some of 
The star clusters studied in this work are obtained by applying our algorithm to the output of the hydrodynamical simulations of  turbulent molecular clouds presented in \cite{ballone20} and \cite{ballone21}. %, where the details on the implementation and the evolution can be found. 
These hydrodynamical simulations are performed with the smoothed-particle hydrodynamics  code {\sc gasoline2} \citep{Wadsley04, Wadsley17}. For this work, we consider the hydrodynamical simulation initialized with a total mass of $2 \times 10^4 \; M_{\odot}$. The cloud has an initial uniform density of 250~cm$^{-3}$, an initial temperature of 10 K and it is in an initial marginally bound state, with a virial ratio $\alpha_{\rm vir}\equiv{} 2\,{}T/|V|=2$, where $T$ and $V$ are the kinetic and potential energy, respectively. The turbulence consists of a divergence-free Gaussian random velocity field, following a \citet{Burgers48} power spectrum. The gas thermodynamics has been treated by adopting an adiabatic equation of state with the addition of radiative cooling \citep{Boley09}. Stellar feedback was not included. Star formation is implemented through a sink particle algorithm adopting the same criteria as in \citet{Bate95}. 

At 3 Myr (for a discussion of this choice see \citealt{ballone21}), we instantaneously remove all the gas from the simulations, mimicking the impact of a supernova explosion. We apply the joining/splitting algorithm to the properties of the sink particles at 3 Myr, as detailed in the next sub-section. 
%, and we apply our joining/splitting algorithm to derive the stellar system with original binaries to be used as initial conditions for \textit{N}$-$body simulations. 
We refer to \cite{ballone20} and \cite{ballone21} for more details on the hydrodynamical simulations.

\subsection{The joining/splitting algorithm} \label{sec_join_split}

 \cite{ballone21} introduced a new algorithm to generate stellar populations from sink particles obtained through hydrodynamical simulations. This algorithm consists in either joining or splitting the sink particle masses, which are affected by non-physical effects (such as the simulation resolution and the adopted sink particle algorithm), so to obtain a new, more realistic mass function of “children” stars. In this way the obtained stellar population inherits the turbulent phase space distribution generated from hydrodynamical simulations, but features a realistic mass function. Here we summarize the main steps of the joining/splitting process. 

First, a  population of stars with a chosen IMF is created, for an assigned value of their total mass. % (the same as in point $\mathrm{a)}$ of Sect. \ref{sec_bin_gen}). 
The joining algorithm is used when a star is more massive than the most massive sink particle. According to the joining algorithm, we select the densest region of the sink particle distribution and merge the neighbour sinks until we obtain the mass of the star. The position and the velocity of the star are assigned as the position and the velocity of the centre of mass of the joined sinks. The joining algorithm tends to enforce mass segregation in the central regions of the simulated star clusters.

The splitting  branch of  the algorithm, instead, is applied if a massive sink is more massive than any left star.  In  this case, we subtract  the mass of individual stars from the massive sink particle, until a mass smaller than 0.1 $\mathrm{M_{\odot}}$ is left. The leftover mass is reassigned to the closest sink, so to enforce local and total  mass  conservation. The children stars of each sink particle are then distributed around the position and velocity of their parent sink according to a virialized Plummer distribution (for this step we make use of the {\sc new\_plummer\_model} module in {\sc amuse}, \citealp{Pelupessy13}). In \cite{ballone21}, we considered a Plummer half-mass radius of $10^{-3}$ pc, that allowed a good energy and virial ratio conservation for all the hydrodynamical simulations of the sample. For this work, we prefer a  Plummer half-mass radius of $10^{-2}$ pc because, for this specific star cluster, this choice allows a better conservation of the total energy and a smaller variation of the virial ratio. 
\begin{comment}
\MP{Non mi è del tutto chiaro (ma forse dipende dal fatto che non ho ben seguito la descrizione nel paper originale) come sostituendo una sink con un Plummer si possa conservare sia la virial ratio che l'energia. Il Plummer è virializzato e se è abbastanza piccolo si comporta come una particella singola nell'interazione colle altre particelle, quindi la virial ratio viene conservata ($U$ nuova = $U$ vecchia + $U$ plummer, $T$ nuova = $T$ vecchia + $T$ plummer) ma allora l'energia non si può conservare perché il plummer introduce una energia totale negativa (è legato; se fosse slegato non sarebbe virializzato). Allora si sceglie $10^{-2}$ perché non sia abbastanza piccolo e il ragionamento di sopra non valga? Oppure si compensa aggiungendo energia cinetica al centro di massa?} \ST{ST: La conservazione di energia e virial ratio dipende da come si combinano joining e splitting in tutta la procedura (vedi anche Fig. 5 di Ballone et al. 2021, anche se la corrispondenza non è perfetta perché per questo lavoro abbiamo cambiato un po' di cose). Per questa simulazione, così come per m1e4, la scelta di $10^{-2}$ rispetto a $10^{-3}$ permette di non alzare troppo l'energia totale gravitazionale. Come conferma se guardi il pannello di sinistra di Fig. 7 di Ballone et al. 2021 puoi notare un aumento di particelle a distanze intorno a $10^{-3}$ pc, dovuto proprio a quella scelta del raggio dei plummer.}
\end{comment}
The process of joining/splitting is cycled until either all the sink particles or the stars are consumed.

\subsection{Direct \textit{N}-body simulations}

For our direct $N-$body simulations, we made use of the direct summation \textit{N}$-$body code {\sc nbody6++gpu} \citep{wang15} coupled with the population synthesis code {\sc mobse} \citep{Mapelli17, giacobbo18,giacobbomapelli18,giacobbo19,mapelli18}, an upgraded version of {\sc bse} \citep{hurley00,hurley02}. 
{\sc nbody6++gpu} implements a 4th-order Hermite integrator, individual block timesteps \citep{makino92} and a Kustaanheimo-Stiefel  regularization of close encounters and few-body subsystems \citep{stiefel65,mikkola93}. A neighbour scheme \citep{nitadori12} is used to compute the force contributions at short time intervals (irregular force/time steps), while at longer time intervals (regular force/time steps) all the members in the system contribute to the force evaluation. The irregular forces are evaluated using CPUs, while the regular forces are computed on GPUs using the CUDA architecture. The force integration includes a solar neighbourhood-like static external tidal field \citep{wang16}. In all our cases, we consider a star as an escaper if it reaches a distance from the centre of density greater than four times the tidal radius of the cluster. The value chosen for the removal distance avoids the presence of potential escapers in the calculation \citep{takahashi12,moyanoloyola13}. {\sc mobse} includes up-to-date prescriptions for massive star winds \citep{giacobbo18}, for core-collapse supernova  explosions \citep{fryer12,giacobbomapelli20} and for pair instability \citep{Mapelli20}. {\sc nbody6++gpu} and {\sc mobse} are  integrated, as described by \cite{dicarlo19,DiCarlo20b}. 

\section{Initial conditions for \textit{N}-body simulations} \label{sec_initial_coditions}

The initial conditions for the \textit{N}$-$body simulations from hydrodynamical simulations (hereafter labeled as \emph{Hydro}) are obtained by combining the binary generation algorithm described in Sect. \ref{sec_bin_gen} and the the joining/splitting procedure (Sect. \ref{sec_join_split}). 
%In particular, we focused on one simulated cluster of the \cite{ballone20} sample, and 
%\micmap{Quello che segue l'ho stroken perch\'e poi lo ripeti meglio} \st{We re-run the selected star cluster under different evolutionary conditions: with and without binary stars and with and without stellar evolution, in order to quantify the impact of binaries and of stellar evolution on its dynamical evolution.}
%We selected the simulation with initial mass of $2 \times 10^4 \; M_{\odot}$, that has been extensively studied in our previous works and presents a dense, well defined central clump where binary interactions with the host environment are expected to be efficient.
%\st{With respect to ballone21, we applied the joining/splitting procedure by considering a different Plummer half-mass radius in the splitting part of the procedure of sink particles. The value selected for the  Plummer half-mass radius is $10^{-2}$ pc (instead of $10^{-3}$ pc) because we evaluated that, for this specific star cluster, this choice allows a better conservation of energy and a smaller variation of the virial ratio.} \ST{(ST: Questa parte l'ho spostata nella sezione sul joining/splitting (segnata in verde).)}
The main properties of the initial conditions for the star cluster are reported in Table~\ref{tab_ic}. The system has a total mass of $M_{\rm tot} = 6687 \; M_{\odot}$, a half-mass radius (defined as the 50\% Lagrangian radius centred in the centre of density)  $r_{50} = 1.70$ pc, and a core radius (defined as the 10\% Lagrangian radius centred in the centre of density) $r_{10}=0.06$ pc. 
After the instantaneous removal of the gas, the system is left in a super-virial state, with $\alpha_{\rm vir}\equiv{}2\,{}T/|V|=1.53$. %{\st{AB: occhio, se alpha= T su V, allora questo numero è la metà. Altrimenti ripristiniamo il fattore 2 nella sezione 2.2, ma per le nubi, allora, 2T su V uguale a  2.}} 

%%%%%%%%TABLE%%%%%%%%%%%
\begin{table}
\begin{center}
\caption{%Properties of the star clusters used as 
Initial conditions of the \textit{N}$-$body simulations.}
\label{tab_ic}
\centering  
\begin{tabular}{l l l l l l}
\hline
%\hline
Name & $M_{\rm tot}$ &  $r_{50}$ & $r_{10}$ & $\alpha_{\rm vir}$ & $f_{\rm bin}$ \\
 & (M$_{\odot}$) & (pc) & (pc) & & \\
\hline
\emph{Hydro} & $6.69 \times 10^3$ & 1.70 & 0.06 & 1.53 & 0.06 \\
\emph{King} & $6.69 \times 10^3$ & 0.42  & 0.06 & 1.53 & 0.06 \\
\emph{Loose Fract} & $6.69 \times 10^3$ & 1.70 & 0.30  & 1.53 & 0.06  \\
\emph{Dense Fract} & $6.69 \times 10^3$ & 0.32 & 0.07 & 1.53 & 0.06  \\
\hline
%\hline
\end{tabular}
\end{center}
\footnotesize{First column: name of the simulation set; second column: total mass $M_{\rm tot}$; third column: half-mass radius $r_{50}$; fourth column: core radius $r_{10}$; fifth column: virial ratio $\alpha_{\rm vir}$; sixth column: binary fraction $f_{\rm bin}$ (if original binaries are present).}
\end{table}
%%%%%%%%%%%%%%%%

In order to quantify %in the most complete way as possible 
the impact of different physical ingredients on the dynamical evolution of a stellar cluster, we take into account four different evolutionary cases:
\begin{itemize}
\item \emph{Bin}: evolution with original binary stars and without stellar evolution.
\item \emph{Bin+SE}: evolution with original binary stars and with stellar evolution. We assumed solar metallicity ($Z=0.02$, \citealt{anders89}), in order to match the young star clusters of the Milky Way \citep{Portegies-Zwart10} and  to maximize the difference with respect to the case without stellar evolution, because mass loss by stellar winds is extremely high at solar metallicity \citep[e.g.,][]{vink01, kudritzki02}. %\AB{\textbf{AB: spiegare, magari in una footnote, perch\'e la differenza \`e massima?}}.
\item \emph{NoBin}: case with no original binary stars and no stellar evolution. 
\item \emph{NoBin+SE}: case without original binary stars but with stellar evolution.
\end{itemize}

The comparison between the aforementioned four different cases allows us to have a complete view of the impact of binaries and of stellar evolution on the dynamical evolution of a cluster with a realistic phase-space distribution of stars. 

For each case, we ran 10 simulations of different joining/splitting realizations in order to filter out stochastic fluctuations.
%%%%%%%%%%%%%%%%%%FIGURE%%%%%%%%%%%%%%%%%%%%%%%%%%%%%%%%%
\begin{figure}
\begin{center}
\includegraphics[width=0.96\hsize, keepaspectratio]{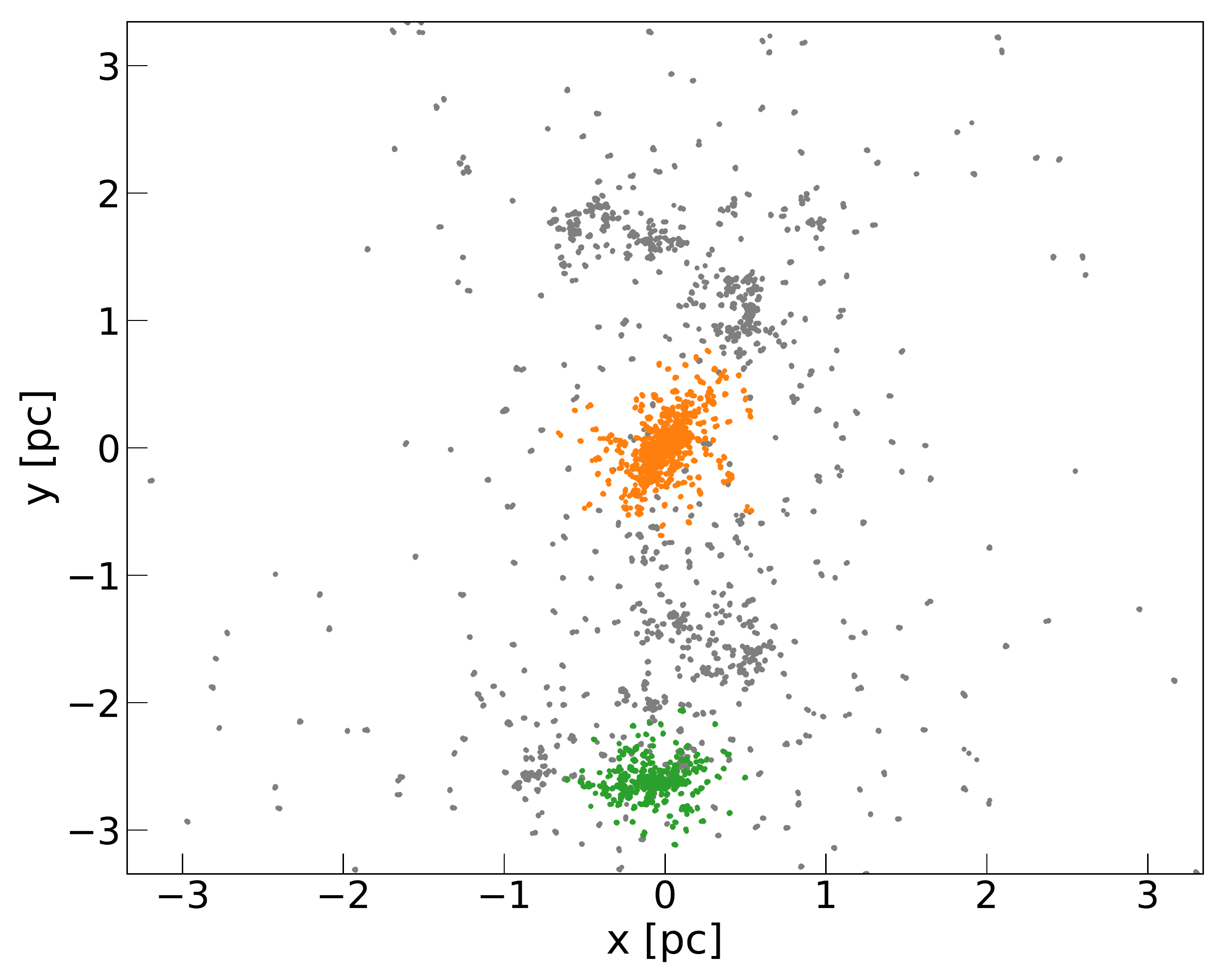}
\includegraphics[width=\hsize, keepaspectratio]{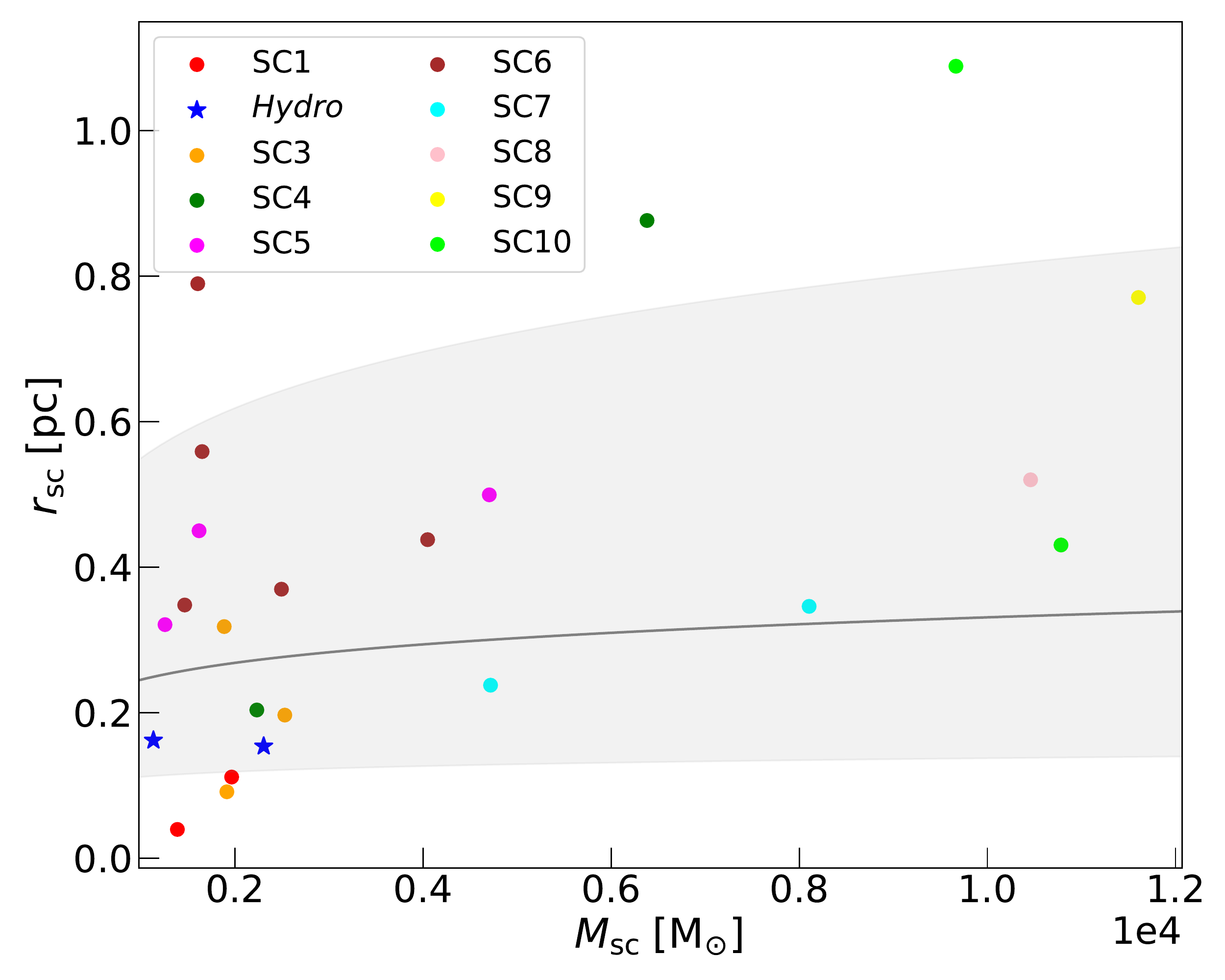}
\caption{\protect{\textbf{Upper panel:}} initial spatial distribution of a realization of the \protect{$Hydro$} stellar cluster after the joining/splitting procedure. The coloured points are the stars that belong to the main (orange) and secondary sub-clump (green). The grey points are the stars that are catalogued as noise points by the \protect{{\sc dbscan}} algorithm. 
\protect{\textbf{Lower panel:}} relation between the mass (\protect{$M_{\mathrm{SC}}$}) and half-mass radius (\protect{$r_{\mathrm{SC}}$}) of all the sub-clumps of the stellar clusters presented in \protect\cite{ballone21}. 
The two sub-clumps of the \protect{$Hydro$} stellar cluster, which corresponds to SC2 in \protect\cite{ballone21}, are marked as blue stars. The grey region is the interval defined by the \protect\cite{MK12} relation (grey solid line). }\label{fig_check_MK}
\end{center}
\end{figure}
%%%%%%%%%%%%%%%%%%FIGURE%%%%%%%%%%%%%%%%%%%%%%%%%%%%%%%%%

%%%%%%%%%%%%%%%%%%FIGURE%%%%%%%%%%%%%%%%%%%%%%%%%%%%%%%%%
\begin{figure*}
\begin{center}
\includegraphics[width=\textwidth]{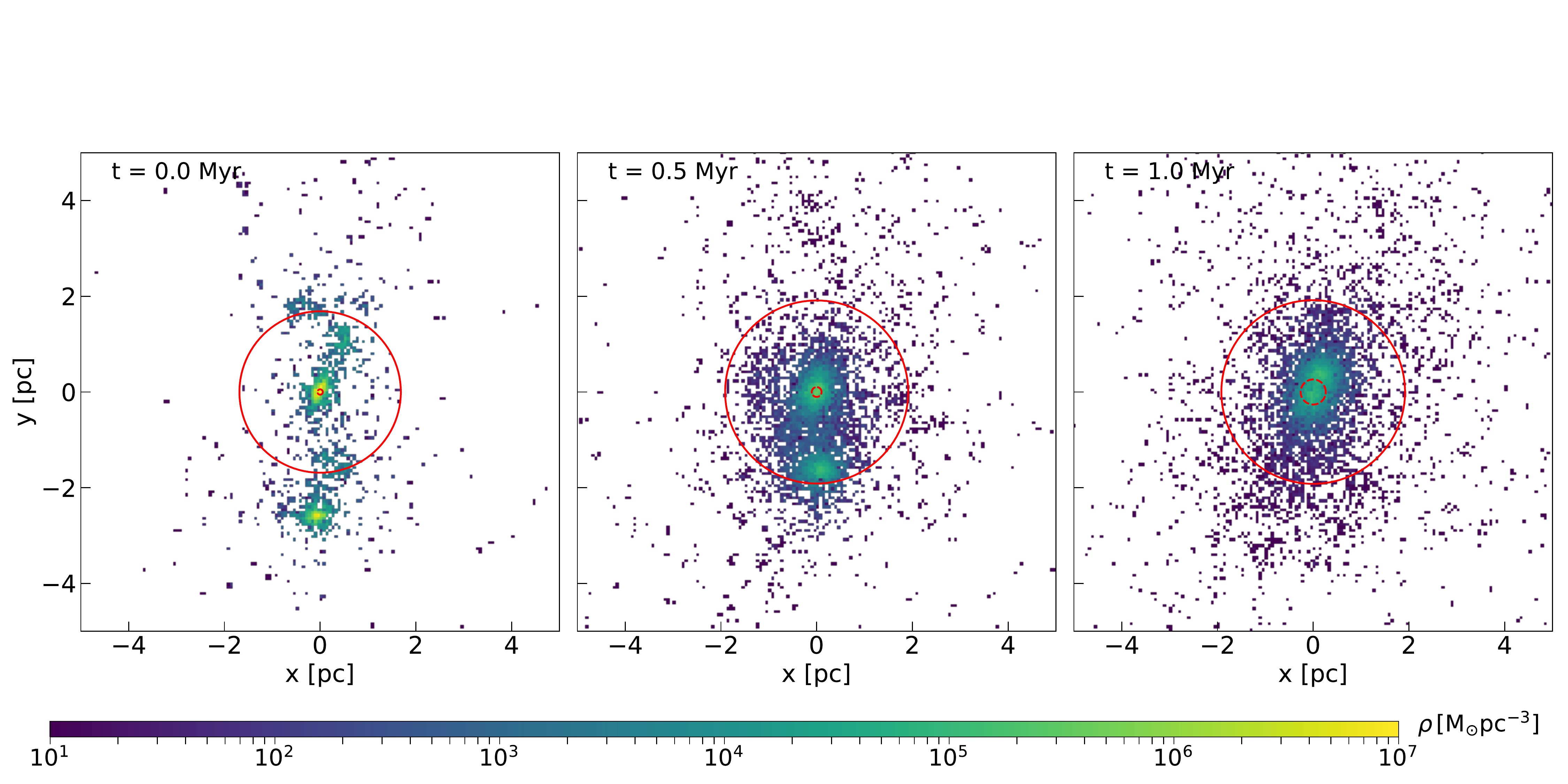}

\caption{Evolution of the cluster  in the first Myr. The red solid line is the half-mass radius, and the red dashed line is the core radius. The left-hand panel shows the initial configuration of the system. The central panel shows the system at 0.5 Myr, when the second sub-clump enters the sphere of the half-mass radius, making it decrease. The right-hand panel shows the system at 1 Myr, when the two main sub-clumps are almost merged and start expanding as a monolithic cluster. Every point is weighted with its local density, calculated as the density of the sphere that includes the 500 closest stars.}\label{fig_movie}
\end{center}
\end{figure*}
%%%%%%%%%%%%%%%%%%FIGURE%%%%%%%%%%%%%%%%%%%%%%%%%%%%%%%%%

\subsection{Comparison with other initial conditions} \label{sec_ic_compariso}

We compared the evolution of the \emph{Hydro} initial conditions to that of %different %(and more traditionally considered) 
other initial phase-space distributions, %of positions and velocities, 
which are commonly used in studies of star cluster dynamics. 
In order to have a fair comparison, we set initial conditions that match the mass scale and either the central length scale ($r_{10}$) or the global length scale ($r_{50}$) of our \emph{Hydro} clusters. All the initial conditions are generated by coupling our binary generation code to {\sc McLuster} \citep{Kupper11} as descibed in Section \ref{sec_bin_gen}. We considered three cases:
\begin{itemize}
    \item \emph{King}: a \cite{king66} model matching the core radius of the hydrodynamical initial conditions. We generated a King model with a half-mass radius of $ r_{50} = 0.25$ pc %(this choice is necessary to obtain the desired core radius), 
    %\st{that roughly corresponds to the size of the central sub-clump}
     and a central concentration of $W_0=9$. The chosen value for the central concentration is high, typical of clusters that are believed to have undergone core-collapse. For this reason a post-core collapse evolution may be expected for both this case and for the central regions of the hydrodynamical case.
    \item \emph{Loose Fract}: a fractal sphere, with the same total mass and half-mass radius as the \emph{Hydro} case. For this case, we selected a fractal dimension $D=1.6$.
    \item \emph{Dense Fract}: a fractal sphere with the same mass and fractal dimension as the previous case, but with a half-mass radius set according to the \cite{MK12} relation:
    \begin{equation} \label{eq_MK}
        r_{50} = 0.10^{+0.07}_{-0.04}\,{} {\rm pc} \left(\frac{M_{\rm tot}}{{\rm M}_{\odot}}\right)^{0.13 \pm 0.04}
    \end{equation}
    %The typical  values obtained for the length scales are 
    In this case, we have $r_{50} \approx 0.3$ pc and  $r_{10} \approx 0.07$ pc. Interestingly, the core radius results very similar to that of the \emph{Hydro} initial conditions.
\end{itemize}
For all these initial conditions we set the same virial ratio as the \emph{Hydro} case. The physical properties for all the initial conditions are summarized in Table~\ref{tab_ic}.

%%%%%%%%%%%%%%%%%%FIGURE%%%%%%%%%%%%%%%%%%%%%%%%%%%%%%%%%
\begin{figure*}
\begin{center}
\includegraphics[width=\textwidth]{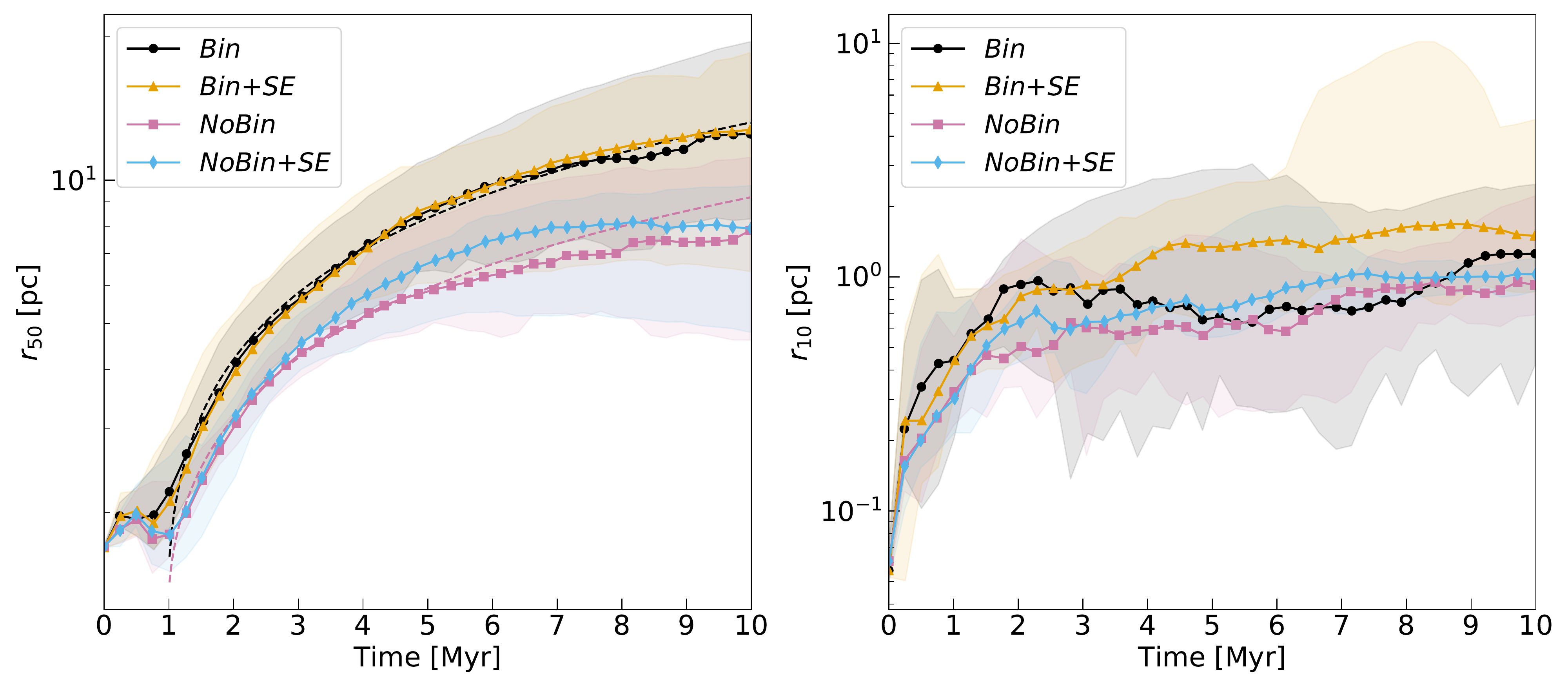}

\caption{Early evolution of the 50\% Lagrangian radius ($r_{50}$, left-hand panel) and 10\% Lagrangian radius ($r_{10}$, right-hand panel) for our set of \textit{N}$-$body simulations. Different lines represent different evolutionary configurations: with original binary stars and without stellar evolution (\emph{Bin}, black circles), with original binary stars and with stellar evolution (\emph{Bin+SE}, ochre triangles), without original  binary stars and without stellar evolution (\emph{NoBin}, pink squares), without original binary stars but with stellar evolution (\emph{NoBin+SE}, cyan diamonds).  For each case, the shaded areas define the range of variation (over the 10 different realizations of each model) of $r_{50}$ (left) and $r_{10}$ (right), while solid lines and markers are the median values. The dashed black and pink lines are our best fit according to eq.~(\protect{\ref{ss_expansion}}).
}
\label{fig_radii}
\end{center}
\end{figure*}
%%%%%%%%%%%%%%%%%%FIGURE%%%%%%%%%%%%%%%%%%%%%%%%%%%%%%%%%

\section{Results}\label{sec_results}

\subsection{Initial clumpiness of the stellar cluster}

The initial space distribution of the $Hydro$ simulation is clumpy and sub-structured, as can be seen in Fig. \ref{fig_check_MK}. The stellar cluster mainly consists of two very dense main sub-clumps and some minor and irregular clusters and filaments. 
We first defined the two main sub-clumps by using the {\sc dbscan} (Density-Based Spatial Clustering of Applications with Noise) algorithm \citep{DBSCAN}\footnote{The implementation we referred to is that of the python library {\sc Scikit-learn} (sklearn.cluster.DBSCAN, \citealp{pedregosa11}). {\sc dbscan} requires to define two parameters, $\epsilon$ and $minPts$. The parameter $minPts$ is the number of points within the reference distance $\epsilon$ needed for a point to be considered as a core point. Otherwise, it is labeled as noise. For our case, we set these parameters based on the half-mass radius of the cluster and on the total number of stars: $minPts = N_{tot}/10$ and $\epsilon=r_{50}/5$.}. This algorithm 
%, \ST{by means of a suitable choice of its parameters $\epsilon$ and $minPts$},
allows to group together points in high-density regions: these are labeled as core points, and they are distinguished from points in low density areas, that are labeled as noise 
\begin{comment}
\MP{Andrebbe detto quale valore di MinPts e di Eps è stato usato. Inoltre i punti che non sono noise (e che quindi appartengono a un cluster) non sono tutti core points, ci sono anche i border points che sono 'density reachable' dai core points ma non sono core points.}. \ST{I parametri iniziali li ho definiti in funzione del raggio di metà massa e del numero totale di stelle per non doverli aggiustare di volta in volta. Per questi cluster $MinPts = N_{tot}/10$ e $eps=r_{50}/5$; l'idea è ovviamente avere un buon riscontro con la distribuzione di massa, non di dare una definizione di subclump con questo algoritmo. Per quanto riguarda i border points, usando sklearn i punti vengono divisi semplicemente in punti di cluster e rumore. E' vero che la classificazione dei border points tra i vari cluster può non essere sempre univoca ma andare nei dettagli in questo caso è davvero troppo...}
\end{comment}
The result of the clustering procedure is shown in the top panel of Fig. \ref{fig_check_MK}: the algorithm manages to identify the two main sub-clusters.

The main sub-clump presents a mass of $M_{\rm sc} \approx{} 2304 \, \mathrm{M_{\odot}}$ (35\% of the total mass) and a half-mass radius of $ r_{\rm SC} = 0.15 \, \mathrm{pc}$, while the second sub-clump has a mass of $M_{\rm sc} \approx{} 1132 \, \mathrm{M_{\odot}}$ (17\% of the total mass) and a half-mass radius of $ r_{\rm SC} = 0.16 \, \mathrm{pc}$. 
We checked if the sub-clump masses and half-mass radii are consistent with eq.~ (\ref{eq_MK}), that is the relation between total mass and half-mass radius found in star-forming cloud cores by \cite{MK12}.  Recently, \cite{fujii21} found that this relation holds in \textit{N}-body/SPH simulations for embedded clusters with mass up to about $10^3 \, \mathrm{M_{\odot}}$ and it is preserved after gas expulsion. In this case, we consider the sub-clumps found in the sample of stellar clusters considered by \cite{ballone21}, that extend to higher masses, between $10^3$ and $10^4 \, \mathrm{M_{\odot}}$. 
As the lower panel of Fig. \ref{fig_check_MK} shows, this sample is well consistent with eq.~(\ref{eq_MK}).

%{\AB{Qua si potrebbe far riferimento ai risultati di Fujii et al. (2001)... Oppure nella discussione, ma non so se riprendi esplicitamente il tema della clumpiness, l\`i...}}

\subsection{Global evolution}
\subsubsection{Early evolution ($t<1 \, \mathrm{Myr}$)}\label{sec_early}

Figure \ref{fig_movie} shows the very first phase ($t\leq{}1 \, \mathrm{Myr}$) of the evolution for one representative cluster. %The red solid circle and the red dashed circle represent $r_{50}$ and $r_{10}$, respectively. 
%Here the local density is calculated as the density ofthe sphere that includes the 500 closest stars.
%\micmap{Qui ti ho tolto io i dettagli inutili ma ricorda per sempre che non devi RI-DESCRIVERE LA CAPTION DELLA FIGURA NEL TESTO.}
At $t=0$ Myr, the centre of density is located well within the main clump, while the second main sub-clump is out of the sphere defined by the half-mass radius.
%The middle panel shows the system at $t=0.5$ Myr. By comparing this snapshot with the initial one, two major mechanisms can be observed.
At $t=0.5$ Myr, the cluster structure has significantly evolved. On the one hand, at small scales, each sub-clump rapidly expands, as a consequence of the instantaneous gas removal, thus lowering its local density. On the other hand, the two main sub-clumps get closer to each other, thus balancing the small scale expansion on a larger scale. These competing mechanisms characterize the first $\approx 1$ Myr of the simulation. 

At $t=1$ Myr the cluster has nearly a monolithic shape. The half-mass radius is slightly larger ($r_{50} \approx 2$ pc) than at the beginning of the simulation (when $r_{50} \approx 1.7$ pc), while the core radius has grown much faster, as can be easily seen from Fig.~\ref{fig_movie}. Typically, a realization reaches a monolithic shape after $1-1.5$~Myr (only in a limited number of cases, this condition is fulfilled at about $2-2.5$~Myr), after a short period in which the two sub-clumps tidally interact without merging. The resultant cluster presents an elongated shape, as a consequence of the strong tidal interaction and the relative motion between the sub-clumps.

The range of merger timescales is in agreement with the results by \cite{fujii15b}, whose simulations can simultaneously reproduce the properties of different types of young star clusters, from massive and dense ones to open clusters and looser OB associations. In this sense, when \textit{N}-body simulations are exploited to study the early evolution of stellar clusters, the timescale of sub-clump mergers is strongly dependent on the initial energetic state of the molecular cloud, as can be inferred by comparing the results in \cite{fujii14} and \cite{ballone21}, who initialized their clouds in a marginally bound state. 
On the observational side, this kind of mergers between sub-clumps seems to be disfavoured to explain the formation of young star clusters like NGC 3603 \citep{banerjee13}, whose observational properties require either a monolithic formation channel or a prompt assembly in $t<1 \, \mathrm{Myr}$. However, the results by \cite{sabbi12} hint that ongoing mergers between very young clusters (such as R136 and the Northeast Clump in NGC 2070) may also occur.

\subsubsection{Cluster expansion}

In order to consider both the initial clumpy evolution and the successive monolithic expansion, we evolved the clusters for 10 Myr. Figure \ref{fig_radii} shows the expansion of the cluster, described by $r_{50}$  and $r_{10}$, for all the four evolutionary cases. 
As a consequence of the mechanism described in Section \ref{sec_early}, the half-mass radius initially grows, reaches a peak at about 0.5 Myr, that is when the secondary sub-clump enters the sphere of the half-mass radius of the main sub-clump, and then decreases. At 1~Myr, $r_{50}$ reaches a minimum and then grows monotonically. The expansion of $r_{50}$ is no longer influenced by the relative sub-clump motion, which at this time have merged or are very close to each other, but is due to the small scale expansion that has now reached larger scales. In contrast, the core radius grows rapidly since the very beginning of the simulation, because the sub-clump motion has no effect at these small scales.

The impact of binary stars is evident in the second phase of the evolution of the cluster, during the monolithic expansion. In fact, clusters with original binaries expand faster after 1 Myr: at this point the large-scale interaction of the sub-clumps is no longer present, and the density in the central region is still high enough (of the order of $10^3-10^4$ M$_{\odot}$ pc$^{-3}$) to allow efficient interactions and energy exchange between the binary stars and their surrounding environment. 
\begin{comment}
\MP{Ok, però anche la velocity dispersion entra nelle sezioni d'urto, l'evoluzione della velocity dispersion potrebbe avere un effetto maggiore su come le binarie interagiscono col resto rispetto all'effetto della densità}.
\end{comment}
In the very first phases, instead, the faster expansion due to binary stars is balanced by the global evolution of the sub-clumps. 

As explained in Sect.~\ref{sec_ic_compariso}, the central regions of the cluster are matched by a King model with $W_0=9$, that is typical of stellar clusters that are thought to have undergone core-collapse. We thus compared the monolithic expansion of the cluster with that expected based on a self-similar evolution, at constant mass \citep{Spitzer87}:
\begin{equation} \label{ss_expansion}
r_{50} = \mathcal{B} \, t^{2/3},    
\end{equation} 
where $\mathcal{B}$ is a proportionality constant. If the evolution of the cluster is a post-core collapse expansion, the time increase of $r_{50}$ should be roughly consistent with eq.~(\ref{ss_expansion}). %\st{even though mass loss is present (it should be $r_{50} \propto t^{(2+\nu)/3}$, with small $\nu$, Spitzer87 Quanto small?).} \ST{(ST: Tolgo quest'ultima frase su $\nu$. Mi sembra di complicare tutto  per niente, visto che alla fine usiamo l'altra formula.)}  
We performed a fit to the median values of $r_{50}$ from 1 Myr curves by using eq.~(\ref{ss_expansion}). The resulting best-fit curves are the dashed lines in the left-hand panel of Fig.~\ref{fig_radii}. We show the curves for the cases \emph{Bin} and \emph{NoBin}, where the lack of stellar evolution should avoid the presence of additional effects (e.g., mass loss) and make the dynamical effect by binaries more evident. 
The curves of both cases seem to be consistent with a post-core collapse phase until 10 Myr.

\begin{figure}
\begin{center}
\includegraphics[width=\hsize, keepaspectratio]{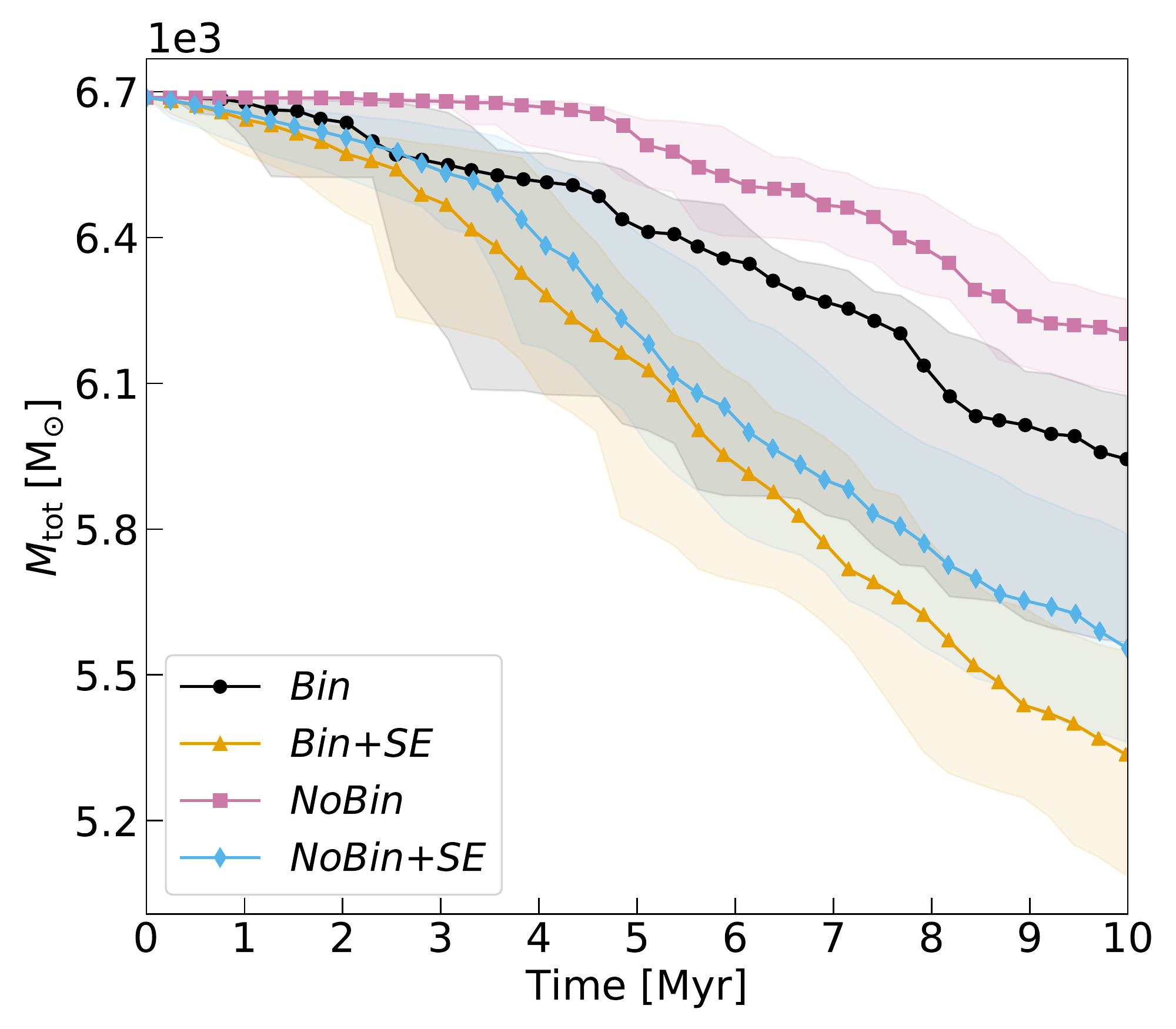}

\caption{Mass variation in the four evolutionary cases we considered. Lines and colours are the same as in Fig.~\ref{fig_radii}. } \label{fig_mass_loss}
\end{center}
\end{figure}
%%%%%%%%%%%%%%%%%%FIGURE%%%%%%%%%%%%%%%%%%%%%%%%%%%%%%%%%

\subsubsection{Mass loss}

As the cluster expands, stars get further away from its centre, until they are eventually removed from the cluster dynamics by the tidal field of the host galaxy. This makes the total mass of the stellar system decrease. The presence of binary stars enhances the number of escaping stars, by powering a faster expansion. Also, close interactions between binary stars and single (or other binary) stars may lead to the ejection of stars, and possibly also of binary systems. In addition, stellar evolutionary processes (e.g. stellar winds, supernova explosions) make single stars, and thus the cluster, lose mass.

%Figure \ref{fig_mass_loss} shows the variation of the total mass of the cluster. Stellar evolution gives the main contribution to mass loss in the early stages of the simulation. In particular, stellar evolution increases the mass loss by about 100\% with respect to the cases when it is not implemented. Also, the only presence of original binaries (Bin) increases the variation of mass by about 60\%  with respect to the case without them (NoBin)

Figure \ref{fig_mass_loss} shows the variation of the total mass of the cluster. Stellar evolution gives the main contribution to mass loss in the early stages of the simulation, resulting in a steeper slope of the mass evolution.  After $10 \, \mathrm{Myr}$, the mass loss in the cases with stellar evolution is twice as large as that in the cases without stellar evolution. 
The absence of primordial binaries delays the mass loss, because the cluster needs to form its binaries dynamically before they start ejecting other stars.
%Also, the presence of primordial binaries adds a shift to the mass loss curves. By comparing the cases \textit{Bin} to \textit{NoBin} and \textit{Bin+SE} to \textit{NoBin+SE} at $10 \, \mathrm{Myr}$, the difference of mass loss due to the only presence of primordial binaries of of about $250 \, \mathrm{M}_{\odot}$. }
%\GN{NG: scusate la domanda propabilemnte stupida ma non mi è chiaro il senso di queste percetuali. Mi sembra che la presenza dell'SE cambi la slope della perdita di massa quindi la percertuale dipende dal tempo in cui si misura la differenza, no?. Invece la binarieta sembra portare ad uno shift rigido (un $\Delta Mtot$ di + 0.1-0.2)}. \ST{ST: No anzi mi sembrano ottime osservazioni. Le ho inserite nel testo.}

%%%%%%%%%%%%%%%%%%FIGURE%%%%%%%%%%%%%%%%%%%%%%%%%%%%%%%%%
\begin{figure*}
\begin{center}
\includegraphics[width=\hsize]{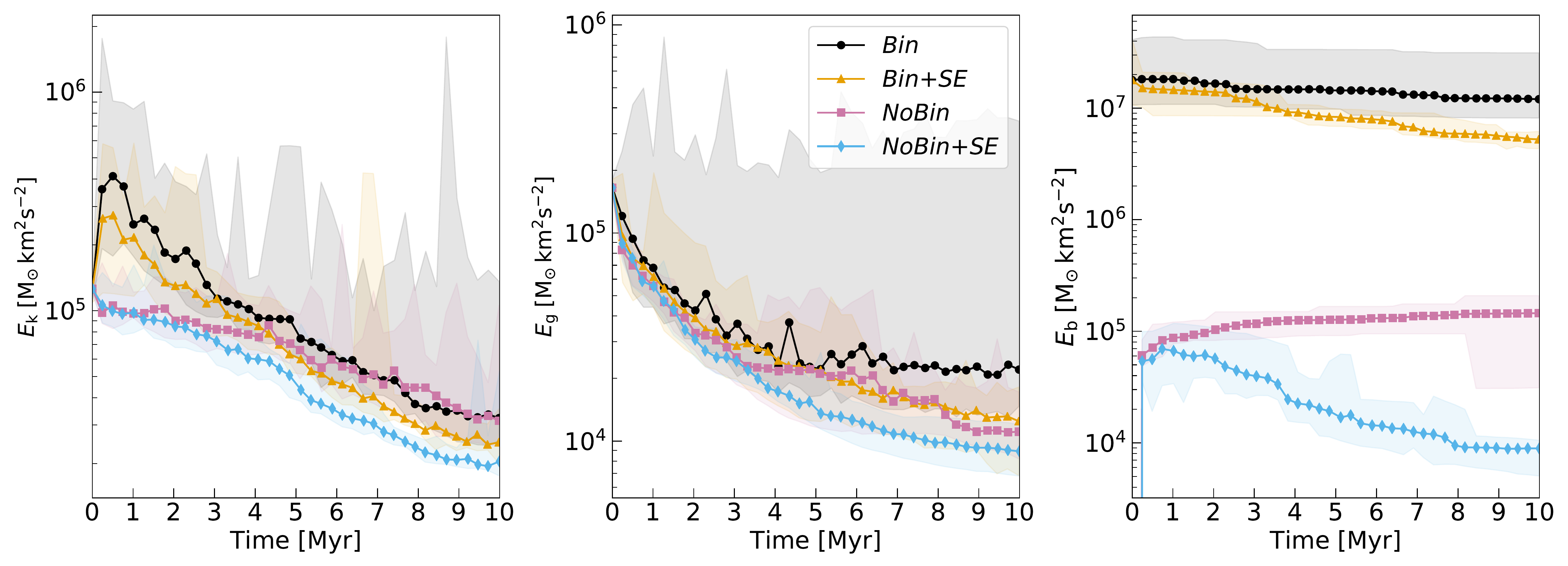}
\caption{Evolution of the total kinetic energy ($E_{\rm k}$, left), of the total gravitational energy  of the centres of mass ($E_{\rm g}$, middle) and of the total binding energy of the binary systems ($E_{\rm b}$, right). Lines and colours are the same as in Fig.~\ref{fig_radii}. 
%\MP{Interessante che la shaded area grigia (binarie senza stellar evolution) sia cosi grande rispetto alle altre. È il range, quindi magari è colpa solo di una simulazione, ma è consistente. E accendere la stellar evolution riduce questa variabilità. Perché?} \ST{Ho controllato e si tratta di due simulazioni (però solo in una l'energia gravitazionale rimane così grande fino alla fine). Sul perché succeda, la prima cosa che mi viene in mente è che si formi un core particolarmente denso che domina l'energia gravitazionale (vedi anche discussione). Forse se c'è stellar evolution, quindi le stelle non sono più punti ma oggetti con raggio finito questo non succede a causa di possibili scontri tra le stelle stesse.}
} \label{fig_energies}
\end{center}
\end{figure*}
%%%%%%%%%%%%%%%%%%FIGURE%%%%%%%%%%%%%%%%%%%%%%%%%%%%%%%%%

\subsubsection{Energy variation}

Figure \ref{fig_energies} shows the evolution of the total kinetic energy ($E_{\mathrm{k}}$), the total  potential energy ($E_{\mathrm{g}}$) of the centres of mass and the total binding energy of binary systems ($E_{\mathrm{b}}$). 
%By considering the kinetic energy, 
Binary stars produce an initial sharp increase of the kinetic energy by yielding their internal energy to the surrounding stars. This results in the fast cluster expansion seen in Fig.~\ref{fig_radii}. %(but in this case it is evident only after the first Myr). 
After this initial sharp increase, the kinetic energy of the clusters with original binaries decreases at a fast rate as a consequence of the ejection or evaporation of high velocity stars. % that rapidly escape from the system. 
%\st{This faster expansion lasts for about 5 Myr, when the total kinetic energies decrease at the same rate.} 
After the first $\sim{}5$ Myr, the kinetic energy of the star clusters with original binary systems becomes similar to that of the other clusters, and they evolve in the same way for the rest of the simulation.

The total binding energy of the initial binary population is much higher than the typical gravitational energy  of the centres of mass. %The binaries of our initial conditions 
Our original binary stars are, in fact, mostly hard and a small fraction of their total internal energy is sufficient to deeply affect the evolution of the cluster. 
The decrease of the total binding energy springs from two factors. Firstly, some binary stars escape from the system. This  causes  the slow decrease of the black line in Fig. \ref{fig_energies}. Secondly, stellar and binary evolution  tend to remove binary stars from the population, via mergers, supernova explosions but also direct collisions between stars. This process is important since the very first stages, because the binary fraction is very high for the most massive stars and because the initial semi-major axes from \cite{sana12} are skewed to small values.  By comparing the \emph{Bin} models  and the \emph{Bin+SE} models, one can infer that this second factor is the main responsible for the variation of the total binding energy.

If there are no original binary systems (\emph{NoBin} and \emph{NoBin+SE} models), the cluster creates its own population, with binding energies of the order of the gravitational energy scale. The case without stellar evolution is characterized by a monotonic increase of the binding energy, where binaries form and the hardest ones tend to harden. In the end, the total binding energy is dominated by the binding energy of a very few binaries. In presence of stellar and binary evolution, after an initial increase, the total binding energy decreases when stellar and binary evolution processes take over.

\subsection{Binary populations}

%%%%%%%%%%%%%%%%%%FIGURE%%%%%%%%%%%%%%%%%%%%%%%%%%%%%%%%%
\begin{figure}
\begin{center}
\includegraphics[width=\hsize, keepaspectratio]{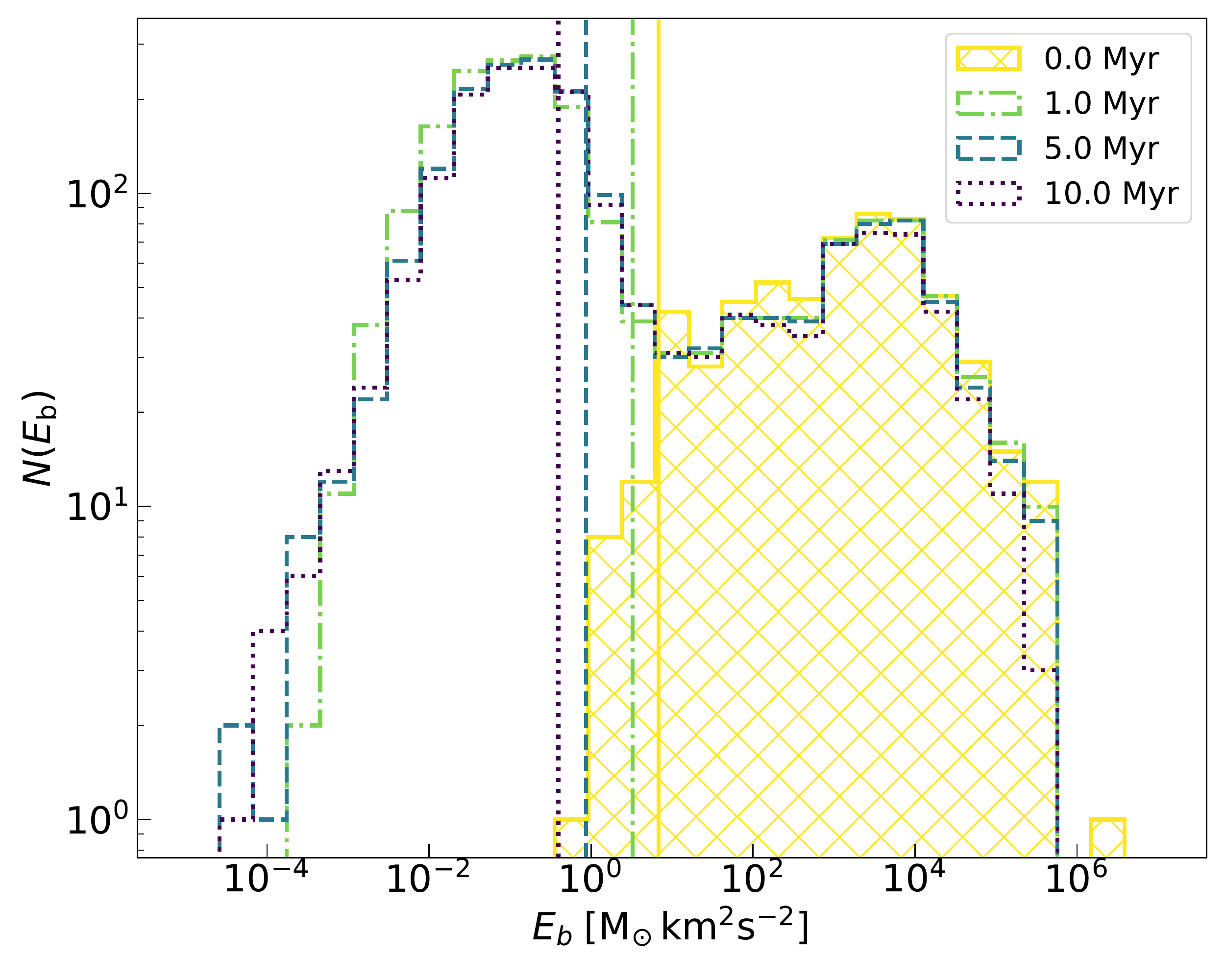}
\caption{Distribution of binding energies for a cluster, in presence of original binaries and stellar evolution. %, in presence of original binaries. 
Four different snapshots are shown: $t=0$ Myr (yellow solid line, hatched area), $t=1$ Myr (green dot-dashed line), $t=5$ Myr (blue dashed line), $t=10$ Myr (purple dotted line). The vertical lines represent the mean kinetic energy of the cluster, defined as the mean kinetic energy of the centres of mass within two half-mass radii (where binaries are more likely to interact).
} \label{fig_binding_energies}
\end{center}
\end{figure}
%%%%%%%%%%%%%%%%%%FIGURE%%%%%%%%%%%%%%%%%%%%%%%%%%%%%%%%%

In order to understand how binary populations evolve and interact with the host cluster, we must estimate how their binding energy distribution is related to the mean energy of the cluster. Figure~\ref{fig_binding_energies} shows the distribution of binding energies for one representative simulation at four different snapshots, in presence of original binary systems and stellar evolution. % (we consider this case to deal with the most realistic framework). 

At the beginning of the simulation, binding energies are very large if compared to the mean kinetic energy. In particular, the hardest part of the distribution is about five orders of magnitude higher than the typical energy scale of the star cluster. This means that the other stars in the cluster "see" the hardest binary systems  as if they were single stars: the cross section of the hardest binary systems is so small that these can hardly interact with single stars.

In absence of original binary systems with a sufficiently large cross section, the star cluster creates new binary systems, with a larger semi-major axis and, thus, a large cross section for three-body encounters. %the cluster is likely to interact with the hardest part of the binary distribution as if they were single stars, and, in absence of interacting binaries, it creates its own.
This is the reason behind the large number of binary systems created at successive snapshots, that are close to the mean kinetic energy of the cluster. Finally, the loosely bound tail of the binary distribution consists of soft binaries that are continuously created and destroyed by dynamical interactions with their neighbours. 

%%%%%%%%%%%%%%%%%%FIGURE%%%%%%%%%%%%%%%%%%%%%%%%%%%%%%%%%
\begin{figure*}
\begin{center}
\includegraphics[width=\textwidth]{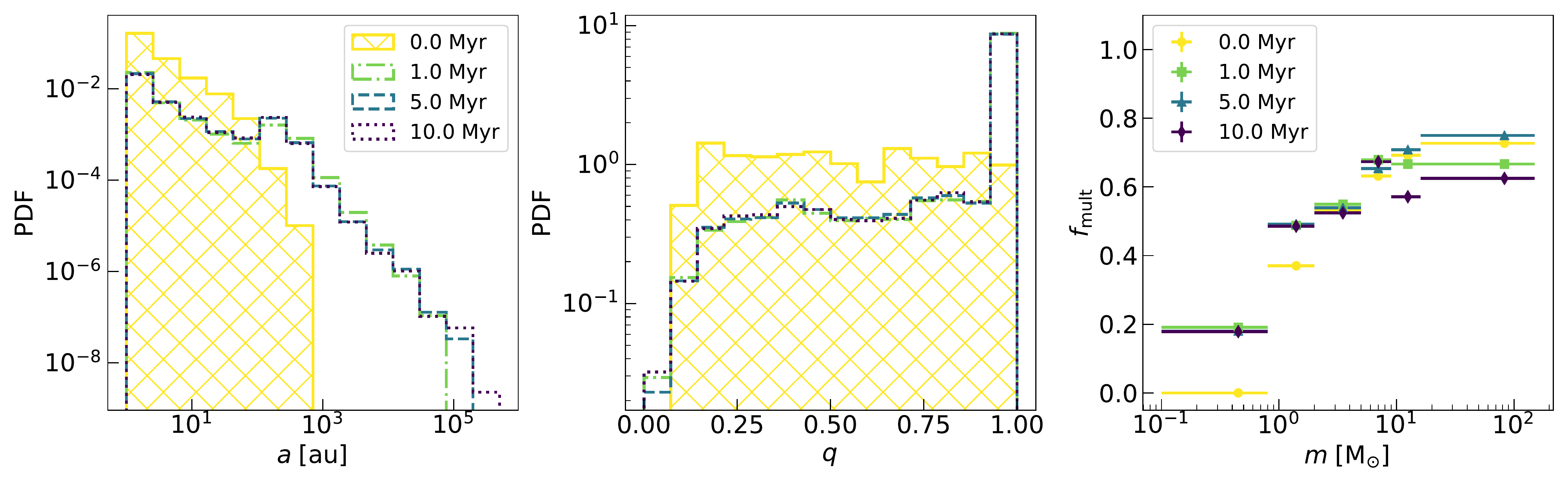}
\includegraphics[width=\textwidth]{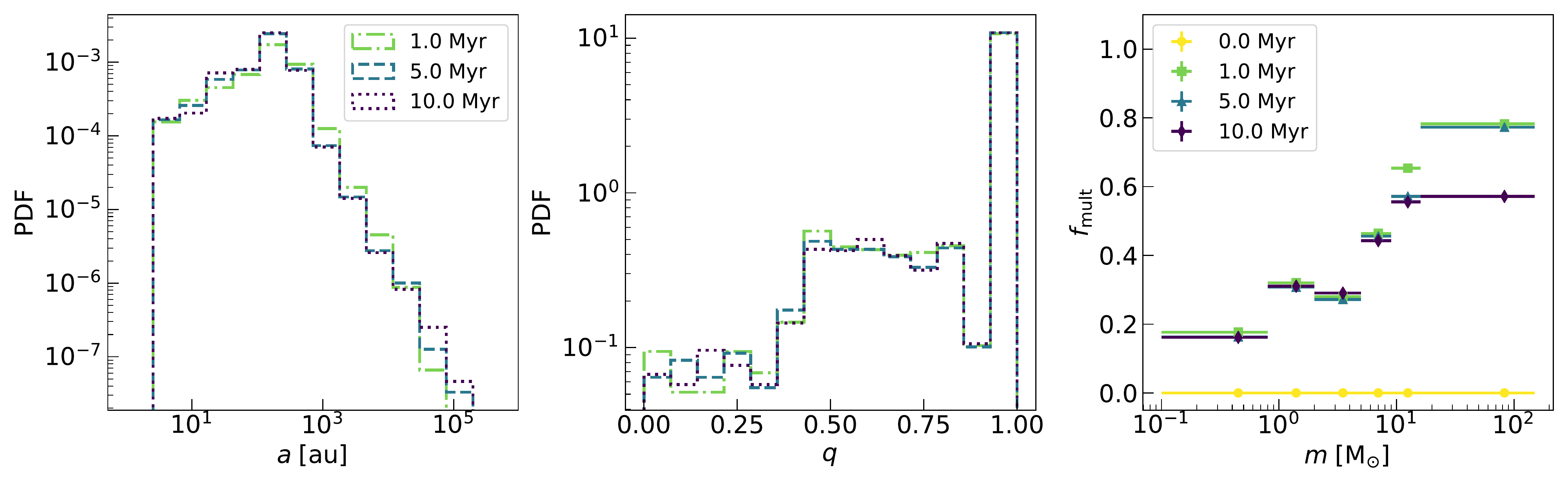}
\caption{Distribution of semi-major axes (left), mass ratios (centre) and multiplicity fractions (right) for a cluster with  (upper panels) and without (lower panels) original binaries. Four different snapshots are shown: $t=0$ Myr (yellow solid line, hatched area, circles), $t=1$ Myr (green dot-dashed line, squares), $t=5$ Myr (blue dashed line, triangles), $t=10$ Myr (purple dotted line, diamonds). }\label{fig_bin_pop_primordial_binaries}
\end{center}
\end{figure*}
%%%%%%%%%%%%%%%%%%FIGURE%%%%%%%%%%%%%%%%%%%%%%%%%%%%%%%%%

\subsubsection{Orbital parameters and multiplicity fraction}

Figure \ref{fig_bin_pop_primordial_binaries} shows the evolution of probability density function (PDF) 
of the binary semi-major axes and of the mass ratios and the multiplicity fraction, defined as the sum of the fraction of binaries and the fraction of triples. 
We consider two representative populations, one for simulations with original binaries (the same as in Fig.~\ref{fig_binding_energies}) and one for simulations without original binaries, in presence of stellar evolution. 

In presence of original binaries, the PDFs significantly change with time, because of  the creation of a large number of dynamical binaries. 
In particular, the distribution of semi-major axes  extends to higher values, and shows a secondary peak at $10^3 $ AU, the typical value at which dynamical binaries form.
This value corresponds to $\approx{}5 \times 10^{-3}$  pc, that is the lowest distance scale (it is the typical distance of stars split into Plummer spheres). As explained above, the cluster responds to the absence of interacting binaries by creating its own. This also explains why the distributions of the dynamically formed semi-major axes and mass ratios are very similar to those that form in absence of original binaries (as shown in the  lower-left panel of Fig.~\ref{fig_bin_pop_primordial_binaries}).  

As for the mass ratios' ($q$) distribution, dynamical interactions produce a steep increase of the PDF at high values, because the new binaries are typically formed by the low mass stars in the Plummer spheres. Also, the distribution of mass ratios extends towards lower values than the initial lower limit ($q=0.1$). Most of the variations in the PDFs take place in the first 1 Myr, that is when the environment is dynamically active. Since then, the binary distributions remain almost unchanged. 
Also, the large number of dynamically-created small-mass binaries  %systems
increases the total multiplicity fraction from $\approx 6 \%$ to $\approx 26 \%$. In particular, these systems populate the lowest mass bin of Figure \ref{fig_binary_distributions}, by increasing the binary fraction from 0 to $20\%$. 

In the absence of original binaries (\emph{NoBin+SE} case), dynamical interactions produce a distribution of semi-major axes that is similar to the distribution of dynamically formed binary systems in the \emph{Bin+SE} case, but cannot reproduce the hardest part of the \cite{sana12} binary distribution. 
Also, dynamical mechanisms tend to create equal-mass binaries. 
Remarkably, the binary fraction of dynamically formed binaries in the \emph{NoBin+SE} case is mass-dependent: it 
%a mass-dependent binary fraction is generated, that 
grows with the mass of the primary star and mimics the trend of the observed distribution \citep{moe17}.
 %\ST{, in agreement with what found by \cite{cournoyercloutier20}}. 
Hence, in the absence of original binary stars, the cluster is able to produce a mass-dependent binary fraction. However, there is not sufficient energy at small scales to reproduce the hardest part of the initial distribution of \cite{sana12}.  

\subsubsection{Exchanges}

\begin{figure}
\begin{center}
\includegraphics[width=\hsize, keepaspectratio]{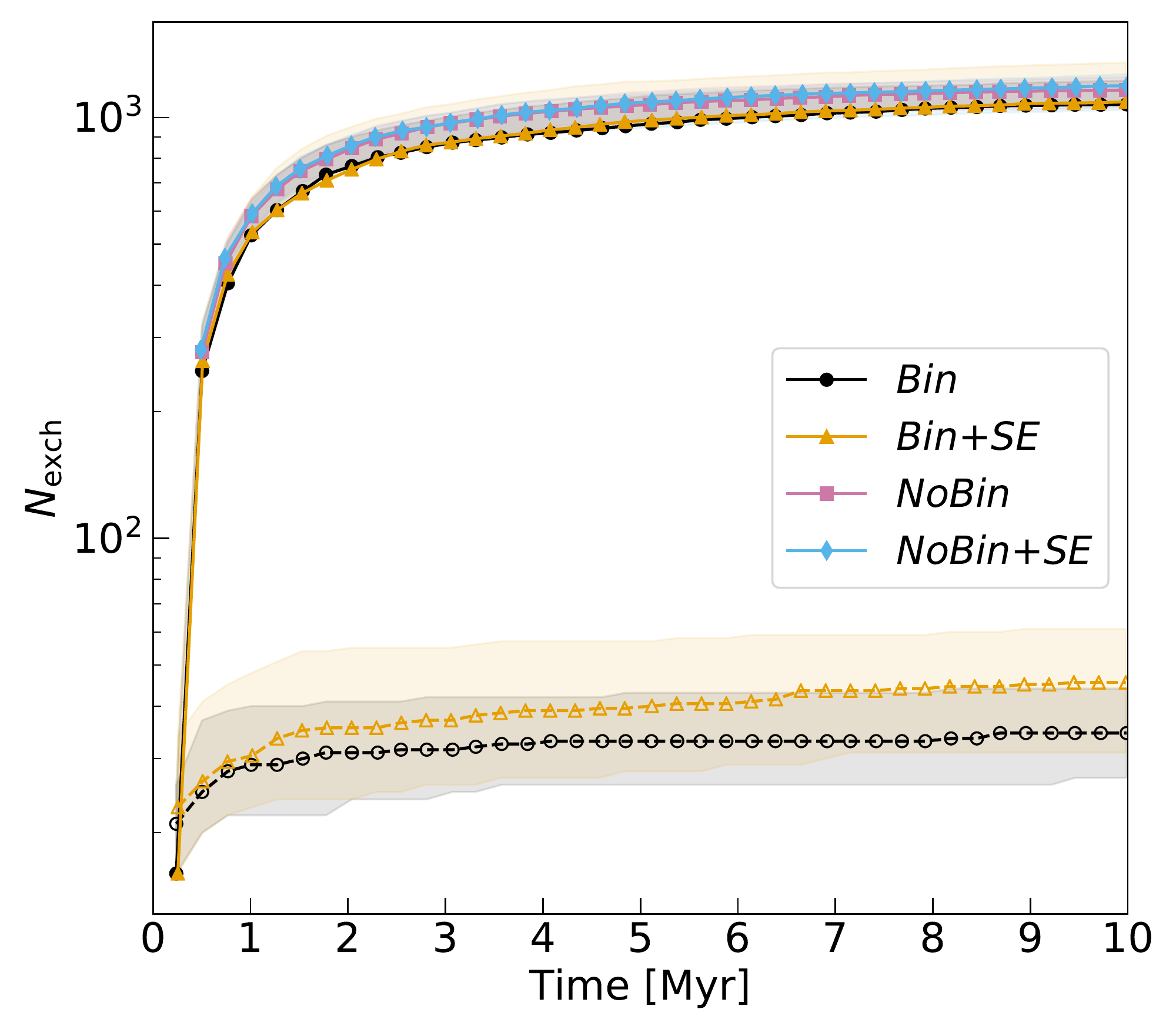}

\caption{Number of exchanges $N_{\rm exch}$ as a function of time for the entire population of binaries (solid lines, filled markers) and for the  sub-population of original binaries (dashed lines, empty markers). $N_{\rm exch}$ is calculated at steps of 0.25 Myr. Lines and colours are the same as in Fig.~\ref{fig_radii}.}\label{fig_exchanges}
\end{center}
\end{figure}

The degree of interactions between the binary systems and their host cluster can be quantified by evaluating the number of exchanges that take place. Fig.~\ref{fig_exchanges} shows the variation of the incremental number of exchanges. %, calculated at a step of 0.25 Myr, is shown. 
%The dashed lines and the hatched areas refer to the original binaries, if there are any. 
The original binaries take part in a  limited number of exchanges, most of which are in the first 2 Myr of the cluster's life, when densities allow an efficient interaction with the other stars. In the following evolution, the original binaries interact much less, as indicated by the flatness of the curve. Nonetheless, because the original binaries are very hard, the few interactions they undergo %allow to 
exchange a sufficient amount of energy to affect the global evolution of the cluster, as shown by the evolution of $r_{50}$ (Fig. \ref{fig_radii}). 

Interestingly, the total number of exchanges is about two orders of magnitude higher than that of original binaries and does not depend on the presence of an initial population of binary stars. This aspect indicates that the cluster under consideration is a very active environment for binary interactions and confirms that the most interacting binaries are %those that are 
dynamically created by the cluster itself. However, most of these exchanges involve binaries that are loosely bound (see also Fig. \ref{fig_binding_energies}) and thus their energy exchange is quite low with respect to that of the original binaries.

%%%%%%%%%%%%%%%%%%FIGURE%%%%%%%%%%%%%%%%%%%%%%%%%%%%%%%%%
%%%%%%%%%%%%%%%%%%FIGURE%%%%%%%%%%%%%%%%%%%%%%%%%%%%%%%%%

%%%%%%%%%%%%%%%%%%FIGURE%%%%%%%%%%%%%%%%%%%%%%%%%%%%%%%%%
%\begin{figure}
%\begin{center}
%\includegraphics[scale=0.25]{Bin+SEm2_e4_4binding_energies10.0.pdf}
%\includegraphics[scale=0.25]{NoBin+SEm2_e4_8binding_energies10.0.pdf}

%\caption{E}\label{radii_set}
%\end{center}
%\end{figure}
%%%%%%%%%%%%%%%%%%FIGURE%%%%%%%%%%%%%%%%%%%%%%%%%%%%%%%%%

\subsection{Comparison with other initial conditions}

The novelty of the \emph{Hydro} initial conditions can be better understood if we compare their evolution to that of other, more idealized initial conditions. To this purpose, we ran simulations with the initial conditions presented in Section~\ref{sec_ic_compariso}. Since we want to focus on the dynamical evolution with different initial phase-space distributions, we decided to run these simulations without stellar evolution.

%%%%%%%%%%%%%%%%%%FIGURE%%%%%%%%%%%%%%%%%%%%%%%%%%%%%%%%%
\begin{figure*}
\begin{center}
\includegraphics[width=\textwidth, keepaspectratio]{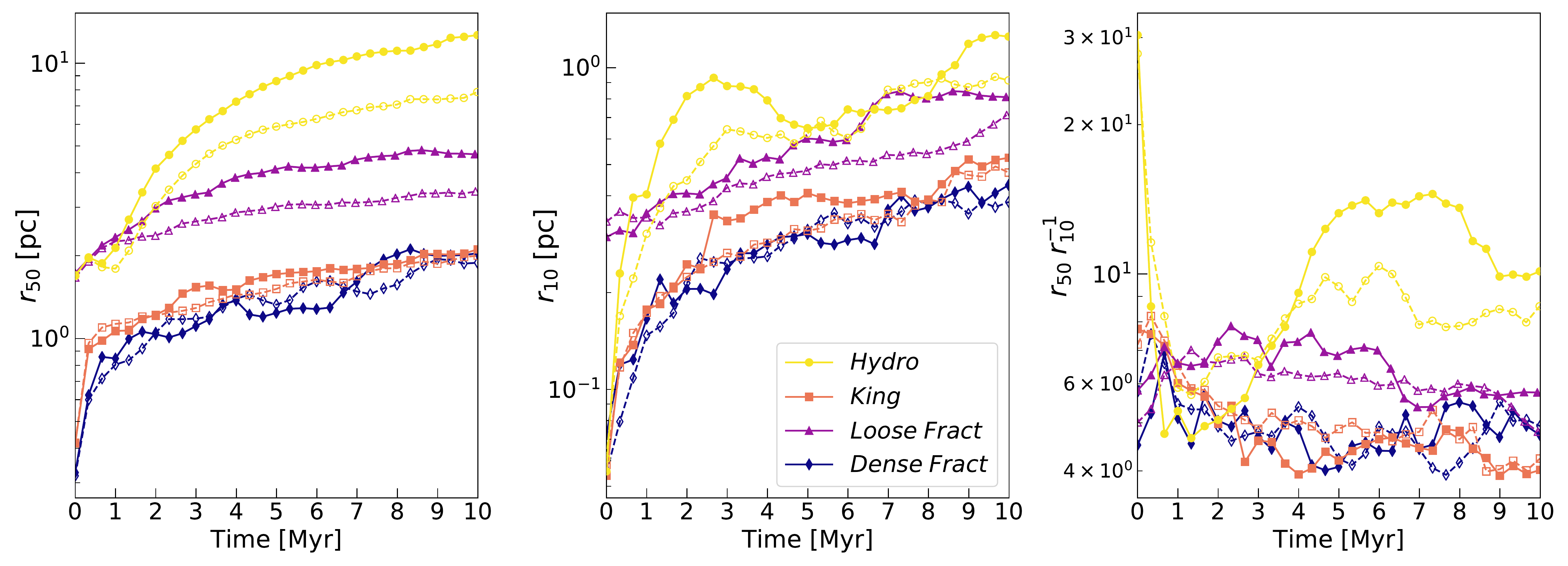}

\caption{Early evolution of the 50\% Lagrangian radius (left-hand panel), the 10\% Lagrangian radius (central panel), and the concentration of the cluster, quantified by $r_{\mathrm{50}}/r_{\mathrm{10}}$ (right-hand panel). Different lines represent the medians of different initial phase-space distributions: \emph{Hydro} (yellow circles), \emph{King} (pink squares), \emph{Loose Fract} (purple triangles), \emph{Dense Fract} (blue diamonds). The solid lines and filled markers represent clusters with original binaries, while the dashed lines and empty markers correspond to clusters without original binaries. 
}\label{fig_radii_comparison}
\end{center}
\end{figure*}
%%%%%%%%%%%%%%%%%%FIGURE%%%%%%%%%%%%%%%%%%%%%%%%%%%%%%%%%

\subsubsection{Cluster Expansion}

Figure \ref{fig_radii_comparison} shows the evolution of the medians of the distributions of $r_{50}$, of $r_{10}$, and of the ratio $r_{50}/r_{10}$, that measures the concentration of the system. In the initial conditions, the \emph{Hydro} clusters have a much larger ratio $r_{50}/r_{10}$ than the other models. Hence, they have very dense cores and rather extended halos, because of the scale of the sub-structures. For these intrinsic differences, the evolution of the characteristic radii of the \emph{Hydro} simulations is considerably different from that of the other distributions.

In the first Myr, the \emph{Hydro} case is the only one that does not show a monotonic increase of $r_{50}$ because of the initial sub-cluster motion (as discussed in Section~\ref{sec_early}). All of the other initial conditions present a monotonic increase of $r_{50}$ and $r_{10}$, but with different slopes. The \emph{Loose Fract} case, that is initialized with the same half-mass radius as the \emph{Hydro} case, shows a mild expansion on both scales, due to its supervirial state. The low density of the central regions (the initial value of $r_{10}$ is larger than in the \emph{Hydro} case by a factor of 5, see Tab.~\ref{tab_ic}) does not allow efficient star-star interactions, that would power a faster expansion. The \emph{King} and  \emph{Dense Fract} models, that are set to match the core radius of the initial \emph{Hydro} simulations, undergo a stronger expansion from the very beginning of their evolution. These two different initial conditions present a very similar behavior. 

The peculiarity in the evolution of the \emph{Hydro} case is evident when the evolution of the ratio $r_{50}/r_{10}$ is taken into account. All the cases except the \emph{Hydro} present a monotonic slow decrease for the $r_{50}/r_{10}$ ratio, that indicates that the systems expand at a similar rate at both scales. 
The  \emph{Hydro} initial conditions, instead, show an initial steep decrease, because the growth of $r_{\mathrm{50}}$ is balanced by the sub-clump motion (see Fig. \ref{fig_radii}), while at smaller scales the cluster expands rapidly. Even when it has reached a nearly monolithic shape, the evolution of its $r_{50}/r_{10}$ ratio is very different from the other initial conditions: this ratio rapidly increases until it reaches a maximum at about 5 Myr. Such a difference may be explained in terms of the stronger mass segregation that features the \emph{Hydro} simulation (we will discuss this point in Sect. \ref{sec_discussion}).

\subsubsection{Binding energies}

%%%%%%%%%%%%%%%%%%FIGURE%%%%%%%%%%%%%%%%%%%%%%%%%%%%%%%%%
\begin{figure}
\begin{center}
\includegraphics[width=\hsize, keepaspectratio]{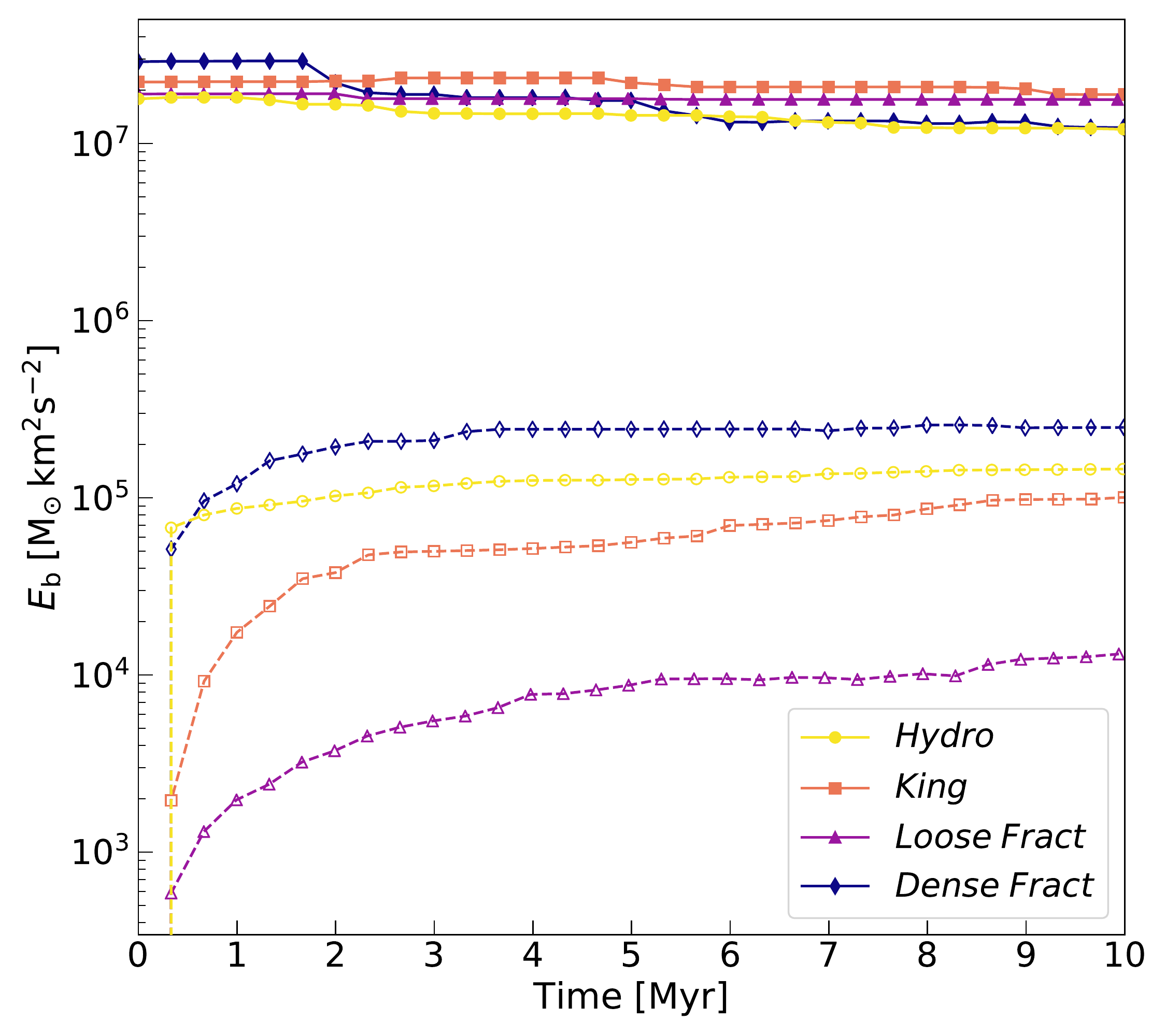}

\caption{Evolution of the total binding energy of the binary systems. Lines and colours are the same as Fig. \ref{fig_radii_comparison}. 
}\label{fig_energies_comparison}
\end{center}
\end{figure}
%%%%%%%%%%%%%%%%%%FIGURE%%%%%%%%%%%%%%%%%%%%%%%%%%%%%%%%%

%In order to highlight the complex interplay between binaries and the host cluster, in 
Figure~\ref{fig_energies_comparison} shows the evolution of the total binding energy for different initial conditions. % in analogy to the right panel of Fig. \ref{fig_energies}. 
In absence of original binaries, every initial configuration creates its own population and the resulting total binding energy is strictly connected to the initial energy scale of the system. 
In particular, the \emph{Hydro}, \emph{King} and \emph{Dense Fract} final binding energies are similar to each other as they are initialized with similar core radii, whereas the total binding energy of the \emph{Loose Fract} systems is about one order of magnitude lower. %\micmap{Questo conferma il mio sospetto sulla concentrazione di cui sopra.}. 
Most of the total binding energy is contained in a  limited number of binaries (from 2 to 5) that go on hardening as the simulation proceeds.    
%\micmap{Qui bisognerebbe citare i lavori di goodman sullo star cluster come oggetto che si autoregola energeticamente.}
This relation between the total binding energy and the global scales of the clusters confirms that star clusters are self-regulating systems  with respect to their binary populations \citep{goodman89,goodman93}: in absence of binaries, each system creates its own population of binaries, with binding energies of the order of its global energy scales.

%\subsection{Stellar Evolution (?)}

\section{Discussion}\label{sec_discussion}

The \emph{Hydro} star clusters present a very distinctive evolution of the $r_{\mathrm{50}}/r_{\mathrm{10}}$ ratio. 
We studied what factor determines the growth of this ratio during the monolithic phase. In particular, we focus on the impact of the initial degree of mass segregation.  
In fact, a high degree of mass segregation would allow the most massive stars to rapidly form a centrally concentrated core that is dynamically separated from the rest of the cluster, the scenario usually referred to as Spitzer instability (\citealp{spitzer69}). If this happens, the distribution of massive stars is hotter than the rest of the cluster, because they remain more concentrated and the local value of the velocity dispersion decreases with the distance from the centre. 

Previous \textit{N}$-$body simulations have found evidence that, for a wide range of initial conditions, the most massive stars in a system do not move slower than the low-mass stars \citep{parker16,spera16, webb17}, as one would expect based on the tendency of stellar systems towards energy equipartition \citep{trenti13,bianchini16}. A confirmation %of the fact 
that the most massive stars can present higher velocities has also been found in proper motion observations of the open cluster NGC 6530 (\citealp{wright19}). \cite{wrightparker19} showed that this aspect can be explained by the combination of Spitzer instability and a cool collapse. If the most massive stars remain more concentrated than the rest of cluster, then the core, that is mostly populated by these massive stars, is expected to expand slower than the rest of the cluster.

\begin{figure}
\begin{center}
\includegraphics[width=\hsize, keepaspectratio]{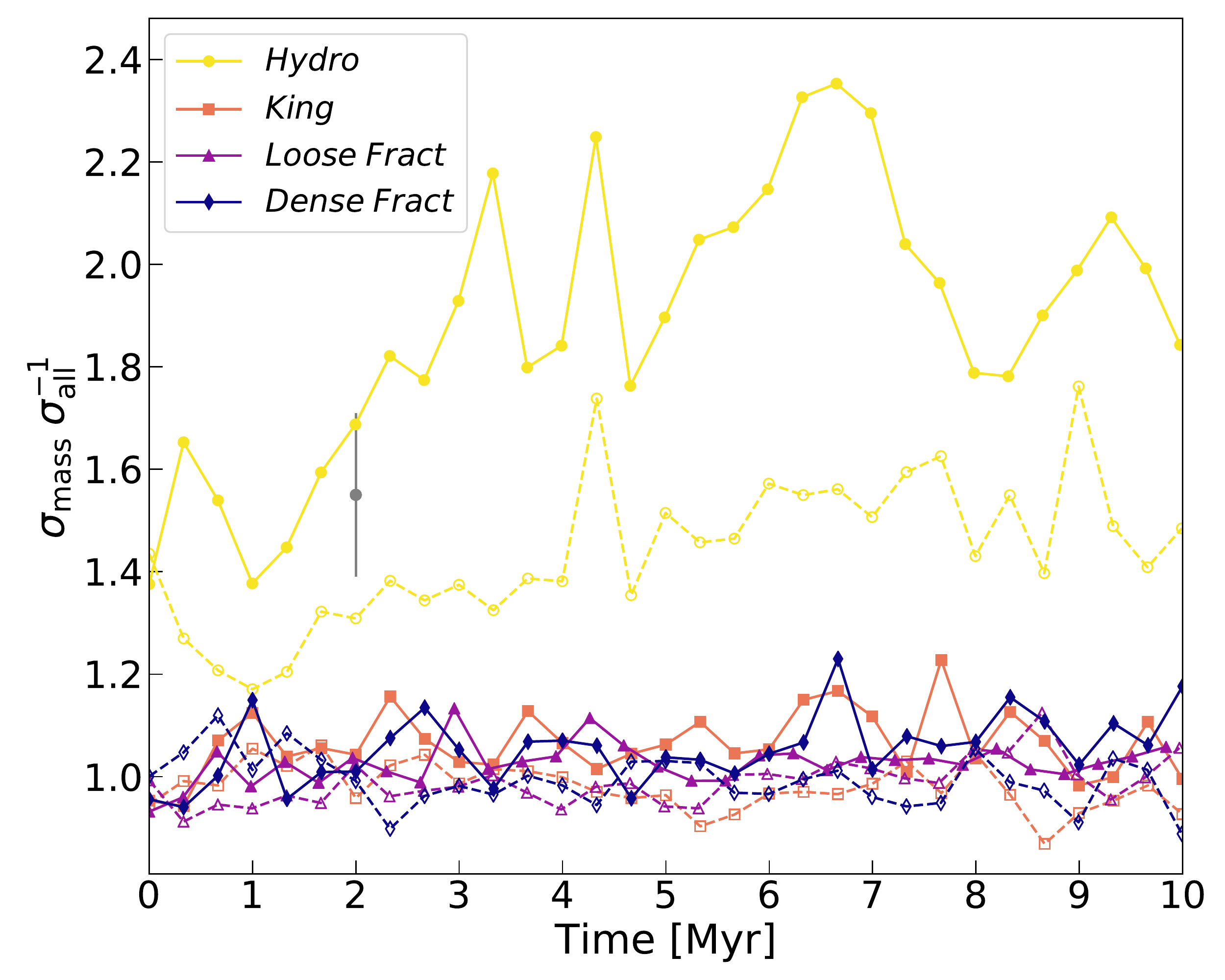}
\includegraphics[width=\hsize, keepaspectratio]{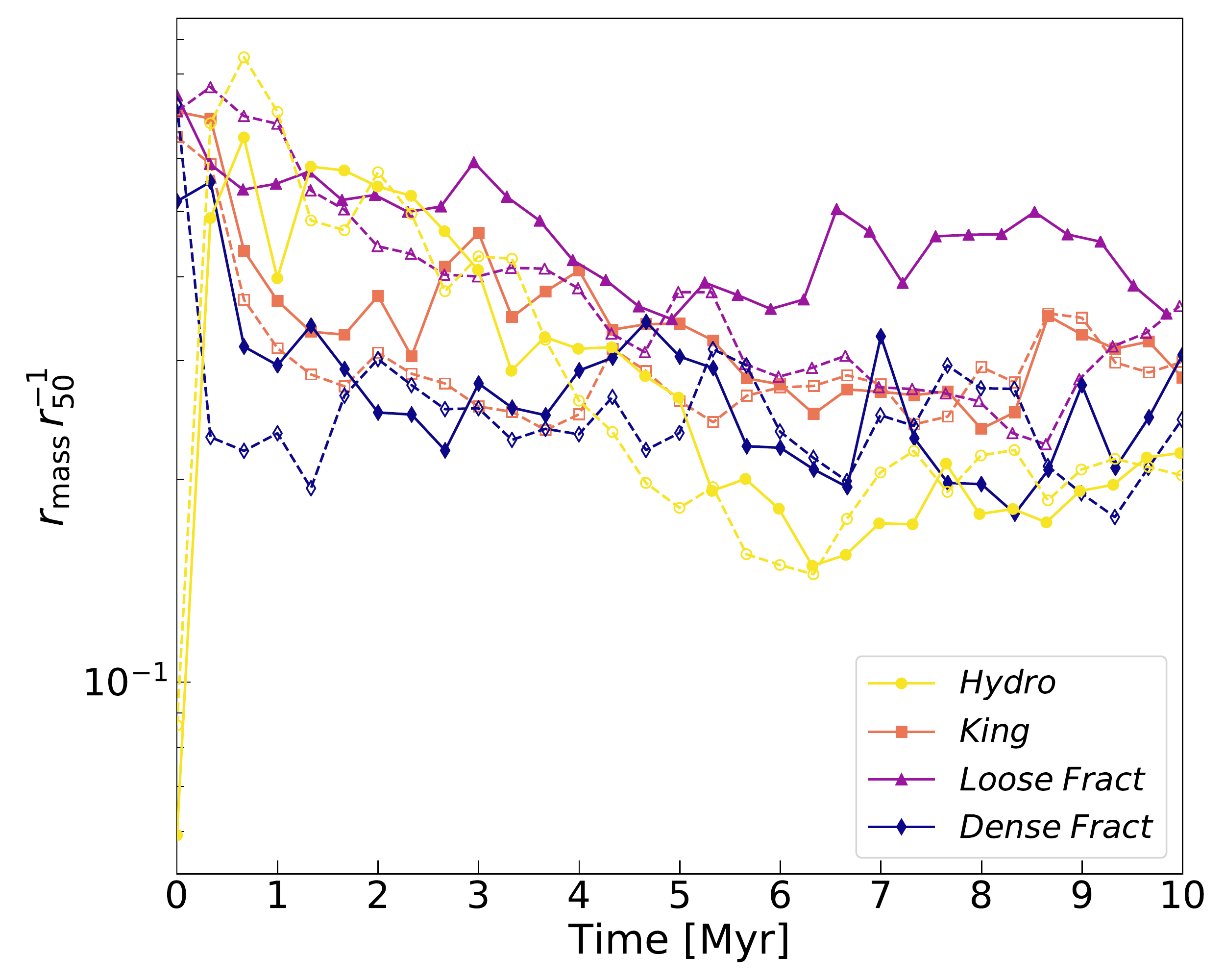}

\caption{\textbf{Upper panel:} evolution of the ratio between the velocity dispersion of the 30 most massive star particles $\sigma_{\rm mass}$ and the velocity dispersion of all the stars inside inside $2 \, r_{50}$, $\sigma_{\rm all}$. The grey data point with error bar is the observed value for NGC~6530 \protect{\citep{wright19}}.  \textbf{Lower panel:} evolution of the ratio between the half-mass radius of the 30 most massive star particles ($r_{\rm mass}$) and the overall half-mass radius, $r_{50}$. Lines and colours are the same as Fig. \ref{fig_radii_comparison}. }\label{fig_mass_segregation}
\end{center}
\end{figure}

To quantify the impact of mass segregation on the evolution of the cluster, we selected the 30 most massive stellar particles\footnote{In the case of a binary, we consider the particle with a mass equal to the total mass of the binary and place in the centre of mass of the binary.} and evaluated the ratio between their velocity dispersion $\sigma_{\mathrm{mass}}$ and the velocity dispersion of all the stellar particles $\sigma_{\mathrm{all}}$. For these calculation, only stars inside two half-mass radii are considered, as done in \cite{wrightparker19}. The evolution of the ratio between these two velocity dispersions is shown in the upper panel Fig.~\ref{fig_mass_segregation}. 
In all the phase-space configurations except the \emph{Hydro}, the velocity dispersion ratio is about one and does not change very much with time. In the \emph{Hydro} case, instead, the high initial value of the velocity dispersion ratio suggests that the stellar cluster presents a strong initial mass segregation. 
Also, during the monolithic phase, the velocity dispersion ratio grows because, after the merger of the two main sub-clumps, their most massive stars rapidly segregate towards the centre, while the system globally expands.  The segregation of the massive stars towards the centre of the potential well may be enhanced by the fact that, in each sub-clump, the stars have already formed a massive core that segregates as one single, very massive particle. In the case with binaries, the velocity dispersion value grows enough to match the observed value for NGC~6530. 
 
The connection between the growth of the velocity dispersion ratio and the degree of mass segregation is confirmed by the trend shown by the ratio of the half-mass radii of the 30 most massive stellar particles $r_{\rm mass}$ and the overall half-mass radius $r_{\rm 50}$, shown in the bottom panel of Fig.~\ref{fig_mass_segregation}. The \emph{Hydro} simulations show an initial strong degree of mass segregation. The initial small scale expansion makes this ratio instantly grow; but, then, it rapidly decreases because of the strong segregation at the centre of the cluster. 
The initial degree of mass segregation seems to be the most important factor in the growth of the velocity dispersion ratio in the \emph{Hydro} case: a stronger initial degree of mass segregation triggers the rapid formation of a dense core that expands more slowly than the rest of the cluster. Also, the rapid formation of a dense core could influences the interaction rate between binaries and the host cluster. If the primordial hard binaries live in a denser environment, they are more likely to interact: %the host environment: 
this explains why the \emph{Hydro} initial conditions present different expansions for the cases with and without binaries (Fig. \ref{fig_radii_comparison}).

%%%%%%%%%%%%%%%%%%FIGURE%%%%%%%%%%%%%%%%%%%%%%%%%%%%%%%%%

\section{Summary and Conclusions} \label{sec_conclusions}

We studied the early dynamical evolution ($t<10 \, \mathrm{Myr}$) of young stellar clusters with realistic populations of binaries and different initial phase-space distributions. 
The initial conditions for our \textit{N}$-$body simulations are obtained by combining a new algorithm to generate realistic stellar and binary distributions (\citealp{sana12, moe17}) with the joining/splitting algorithm defined in \cite{ballone21}, to derive initial conditions from hydrodynamical simulations. %, and the {\scshape McLuster} \cite{Kupper11} code, to obtain spherical initial conditions.   

For the hydrodynamical initial conditions (\emph{Hydro} cluster), we considered different evolutionary cases by switching on and off the presence of original binary stars and stellar evolution in order to weight their contribution to the dynamical evolution.
Our results show that the evolution of the cluster is characterized by two distinct evolutionary phases: first, the global expansion of the cluster is balanced by the approaching of its main sub-clumps, while at small scales the cluster expands instantaneously. After 1 Myr, the cluster has reached a nearly monolithic shape and expands as a whole, following a post-core collapse expansion. Binaries tend to speed up the expansion of the cluster in this phase, making the half-mass radius expand faster, while stellar evolution has a minor impact on the early dynamical evolution of the cluster, but has a major impact on  mass loss. 

We compared the evolution of the \emph{Hydro} star cluster to that of star clusters with spherical distributions of stars (\emph{King}, \emph{Loose Fract}, \emph{Dense Fract}). The main difference between the \emph{Hydro} cluster and the others relies in the evolution of the $r_{50}/r_{10}$ ratio, that measures the concentration of the system. The \emph{Hydro} cluster, in fact, shows a distinctive trend of $r_{50}/r_{10}$. At the beginning of the simulations, $r_{50}/r_{10}$ is much larger in the \emph{Hydro} cluster than in the other models, because the \emph{Hydro} cluster is an aggregate of several sub-clumps, resulting in a large total half-mass radius, but its core radius is very small, since it basically coincides with the core radius of the densest sub-clump. The $r_{50}/r_{10}$ ratio decreases very fast ($<1$ Myr) in the \emph{Hydro} cluster, reaching values similar to the other clusters, because of the hierarchical merger of the sub-clumps, which reduces the total half-mass radius. However, at $t>1$ Myr the value of $r_{50}/r_{10}$ keeps decreasing in the spherical models, while it grows again in the \emph{Hydro} case. The late growth of $r_{50}/r_{10}$ in the \emph{Hydro} cluster is due to its initial high degree of mass segregation, which allows it to form a centrally concentrated core of massive stars. As this core expands more slowly than the rest of the cluster, the ratio between the velocity dispersion of the most massive stars and that of all the stars increases. In the case with binaries, it grows enough to match the observed value for NGC~6530 \citep{wright19}. 

The initial %population of 
binary stars we set based on observational constraints \citep{sana12,moe17} are  generally too hard to interact in an efficient way with the host environment. The stellar systems recover from the lack of interacting binaries by dynamically creating additional binaries with binding energy of the order of their kinetic energy. Also, in the absence of primordial binaries, the binary fraction of the dynamically formed binaries show a binary fraction that increases with the mass of the primary star. This behaviour spontaneously reproduces the relation between binary fraction and stellar mass found in observations \citep{moe17}. %mimics the observed trend \citep{moe17}.

%{\AB{AB: Penso che qualche riferimento alle crtiche di Kroupa ci voglia, magari quando rispieghiamo la early evolution ed il merger dei subclump. Forse dovremmo anche trovare il modo di citare gli ultimi lavori dell Fujii...}} \ST{ST: Ho aggiunto in Sezione 4.1 e 4.2.}

\section*{Acknowledgements}

MM, AB, UNDC, NG and SR acknowledge financial support by the European Research Council for the ERC Consolidator grant DEMOBLACK, under
contract no. 770017. NG is supported by Leverhulme Trust Grant No. RPG-2019-350 and Royal Society Grant No. RGS-R2-202004. MP's contribution to this material is supported by Tamkeen under the NYU Abu Dhabi Research Institute grant CAP3. We acknowledge the CINECA-INFN agreement, for the availability of high performance computing resources and support. We also thank Simon Portegies Zwart for useful discussions.

%%%%%%%%%%%%%%%%%%%%%%%%%%%%%%%%%%%%%%%%%%%%%%%%%%

\section*{Data availability}
The data underlying this article will be shared on reasonable request to the corresponding authors.

%%%%%%%%%%%%%%%%%%%% REFERENCES %%%%%%%%%%%%%%%%%%

% The best way to enter references is to use BibTeX:

\bibliographystyle{mnras}
\bibliography{bibliography} % if your bibtex file is called example.bib

%% Alternatively you could enter them by hand, like this:
%% This method is tedious and prone to error if you have lots of references
%\begin{thebibliography}{99}
%\bibitem[\protect\citeauthoryear{Author}{2012}]{Author2012}
%Author A.~N., 2013, Journal of Improbable Astronomy, 1, 1
%\bibitem[\protect\citeauthoryear{Others}{2013}]{Others2013}
%Others S., 2012, Journal of Interesting Stuff, 17, 198
%\end{thebibliography}

%%%%%%%%%%%%%%%%%%%%%%%%%%%%%%%%%%%%%%%%%%%%%%%%%%

%%%%%%%%%%%%%%%%% APPENDICES %%%%%%%%%%%%%%%%%%%%%

\begin{comment}
\appendix
\end{comment}
%%%%%%%%%%%%%%%%%%%%%%%%%%%%%%%%%%%%%%%%%%%%%%%%%%

% Don't change these lines
\bsp	% typesetting comment
\label{lastpage}
\end{document}